\documentclass[journal]{IEEEtran}
\usepackage{algorithm, algorithmic}
\usepackage{graphicx}
\usepackage{epstopdf, xspace}
\usepackage{amsmath}
\usepackage{color}
\usepackage{ifthen}
\usepackage{subcaption}
\usepackage[normalem]{ulem}
\usepackage{amssymb}
\usepackage[font=small,labelfont=bf]{caption}
\usepackage{amsthm, times, amsfonts, cite}

\newtheorem{theorem}{Theorem}
\newtheorem{lemma}{Lemma}
\newtheorem{proposition}{Proposition}
\newenvironment{thisnote}{\par}{\par}

\DeclareCaptionLabelFormat{andtable}{#1~#2  \&  \tablename~\thetable}

\newcommand{\etc}{\textit{etc.}\xspace}
\newcommand{\ie}{\textit{i.e.,}\xspace}
\newcommand{\eg}{\textit{e.g.,}\xspace}

\newcommand{\final}{}

\begin{document}
\definecolor{brown}{cmyk}{0,0.81,1,0.60}
\definecolor{magenta}{rgb}{0.4,0.7,0}
\definecolor{gray}{rgb}{0.5,0.5,0.5}
\definecolor{red}{rgb}{1,0,0}
\definecolor{green}{rgb}{0.5,0,0.5}
\definecolor{blue}{rgb}{0,0,1}


\ifthenelse{\isundefined{\final}} {
\newcommand{\vaneet}[1]{{\color{green}(VA: #1)}}
\newcommand{\shuai}[1]{{\color{blue}(SH: #1)}}
\newcommand{\feng}[1]{{\color{red}(FQ: #1)}}
\newcommand{\anis}[1]{{\color{brown}(AE: #1)}}
\newcommand{\shubho}[1]{{\color{magenta}(SS: #1)}}
}{
\newcommand{\vaneet}[1]{}
\newcommand{\shuai}[1]{}
\newcommand{\feng}[1]{}
\newcommand{\anis}[1]{}
\newcommand{\shubho}[1]{}
}

\newcommand{\eat}[1]{}

\newcommand{\BULLET}{\vspace{+.03in} \noindent $\bullet$ \hspace{+.00in}}

\title{LBP: Robust Rate Adaptation Algorithm for SVC Video Streaming}
\author{Anis Elgabli, Vaneet Aggarwal, Shuai Hao, Feng Qian, and Subhabrata Sen\thanks{A. Elgabli and V. Aggarwal are with Purdue University, West Lafayette IN 47907 (email: aelgabli@purdue.edu, vaneet@purdue.edu). S. Hao and S. Sen are with AT\&T Labs-Research, Bedminster NJ 07921 (\{haos, sen\}@research.att.com). F. Qian is with Indiana University, Bloomington IN 47405 (email: fengqian@indiana.edu). The work of A. Elgabli and V. Aggarwal was supported in part by the U.S. National Science Foundation
under grants CCF-1527486 and CNS-1618335.
}}

\maketitle

\begin{abstract}
	Video streaming today accounts for up to 55\% of mobile traffic. In this paper, we explore streaming videos encoded using Scalable Video Coding scheme (SVC) over highly variable bandwidth conditions such as cellular networks. SVC's unique encoding scheme allows the quality of a video chunk to change incrementally, making it more flexible and adaptive to challenging network conditions compared to other encoding schemes. Our contribution is threefold. First, we formulate the quality decisions of video chunks constrained by the available bandwidth, the playback buffer, and the chunk deadlines as an optimization problem. The objective is to optimize a novel QoE metric that models a combination of the three objectives of minimizing the stall/skip duration of the video, maximizing the playback quality of every chunk, and minimizing the number of quality switches. Second, we develop Layered Bin Packing  (LBP) Adaptation Algorithm, a novel algorithm that solves the proposed optimization problem. Moreover, we show that LBP achieves the optimal solution of the proposed optimization problem with linear complexity in the number of video chunks. Third, we propose an online algorithm (online LBP) where several challenges are addressed including handling bandwidth prediction errors, and short prediction duration. Extensive simulations with real bandwidth traces of public datasets reveal the robustness of our scheme and demonstrate its significant performance improvement as compared to the state-of-the-art SVC streaming algorithms. The proposed algorithm is also implemented on a TCP/IP emulation test bed with real LTE bandwidth traces, and the emulation confirms the simulation results and validates that the algorithm can be implemented and deployed on today's mobile devices. 
\end{abstract}



\section{Introduction}



Mobile video has emerged as a dominant contributor to cellular traffic. It already accounts for around $40-55$ percent of all cellular traffic and is forecast to grow by around 55 percent annually through 2021~\cite{ericsson_report}.
While its popularity is on the rise, delivering high quality  streaming video over cellular networks  remains extremely challenging. In particular, the video quality under challenging conditions such as mobility and poor wireless channel is sometimes unacceptably poor. Almost every viewer at some point in time can relate to  experiences of choppy videos, stalls, \etc


Not surprisingly, a lot of attention from both research and industry in the past decade has  focused on the development of  {\em adaptive} streaming techniques for video on demand that can dynamically adjust the quality of the video being streamed to the  changes in network conditions.  Such a scheme has 2 main components:

\BULLET {\em Content Encoding}: On the server side, the video is divided into multiple \emph{chunks} (segments), each containing data corresponding to some playback time  (e.g., 4 sec), and then each chunk is encoded  at multiple resolutions/quality levels~(each with different bandwidth requirements).

\BULLET {\em Adaptive Playback}: During playtime, an entity (typically the player) dynamically switches between the different available quality levels  as it requests the video over the network. The adaptation is based on many factors such as the network condition, its variability, and the client buffer occupancy \etc. This results in a viewing experience where different chunks of the video might be streamed at different quality levels.

In the predominant adaptive coding technique in use today, each video chunk is stored into $L$ {\em independent} encoding versions, as an example of such a technique is H.264/MPEG-4 AVC (Advanced Video Coding) which was standardized in 2003~\cite{H264}.
%
During playback when fetching
a chunk, 
the Adaptive Bit Rate (ABR) streaming technique such as MPEG-DASH~\cite{DASH} (Distributed Dynamic Streaming over HTTP) needs to select one out of the $L$ versions based on its judgement of the network condition and other aforementioned factors.

\begin{figure}[t]
   \centering
   \small
   \includegraphics[width=.25\textwidth]{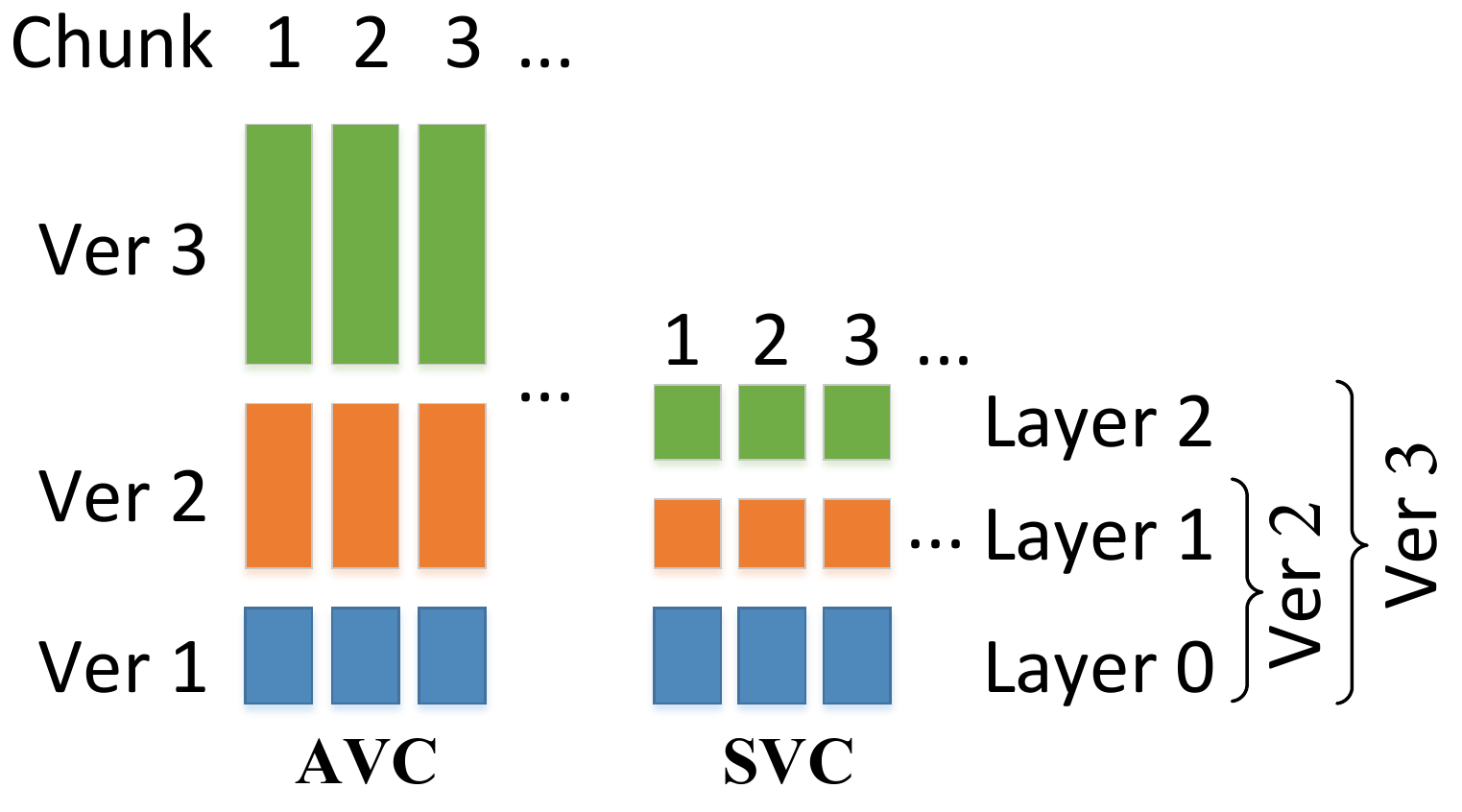}
   \caption{AVC vs SVC Encoding}
   \label{fig:encoding}
   \vspace{-.in}
\end{figure}

An alternative encoding scheme is Scalable Video Coding (SVC) which was standardized in 2007 as an extension to H.264~\cite{4317636}. In SVC, a chunk is encoded into ordered \emph{layers}: one \emph{base layer} (Layer 0) with the lowest playable quality, and multiple \emph{enhancement layers} (Layer $i>$0) that further improve the chunk quality based on layer $i-1$. When downloading a chunk, an Adaptive-SVC streaming logic must consider fetching all layers from 0 to $i-1$ if layer $i$ is decided to be fetched. In contrast, in AVC, different versions (\ie qualities) of chunks are independent, as illustrated in Fig.~\ref{fig:encoding}.

%


 There are  three typical modes of scalability, namely temporal (frame rate), spatial (spatial resolution), and quality (fidelity, or signal-to-noise ratio). The encoding however has an additional encoding overhead, which depends on the mode of scalability.  For example \cite{mathew2010overview} showed that there is minimal or no loss in coding efficiency using temporal scalability. Temporal scalability is also backward compatible with existing H.264 decoders, and is simple to implement as compared to other forms of scalability. However, there are some limitations for using temporal scalability such as being visually un-pleasing for low base layer rates which motivate the use of other scalability modes that require more overhead. 
  Appendix \ref{sec:whySVC} describes some common scenarios where Adaptive-SVC streaming can be beneficial.

\begin{figure}
	\vspace{-.1in}
	\includegraphics[trim=0in 1.6in .5in 1.5in, clip,width=.48\textwidth]{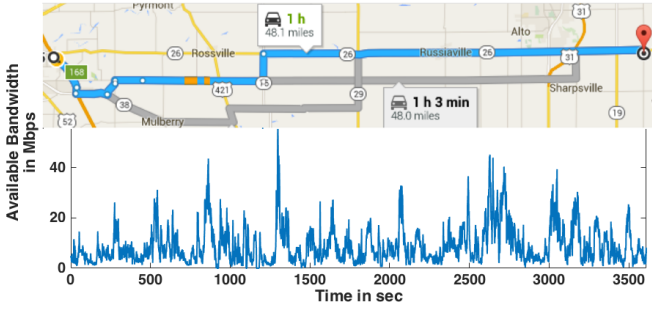}
	\caption{Motivating example: network condition prediction can improve streaming quality under mobility.}
	\label{fig:route}
	\vspace{-.1in}
\end{figure}

To motivate our problem, imagine a scenario where a mobile user starts a trip from point A to point B (see Fig. \ref{fig:route}, anonymized with randomly chosen locations). As the user enters the destination location to the GPS application, she gets the route information, and the video player obtains the estimates of the bandwidth availability along the chosen path.  The bandwidth estimation can be obtained using crowd-sourced information from measurements of other users who travelled the same route recently as we will show in Appendix \ref{sec:predict}.
We will demonstrate that access to such information can help the player take significantly better informed decisions in its adaptation logic. For example, if the player is aware that it is about to traverse through a region with low bandwidth, it can switch to fetching the video at a lower quality to minimize the possibility of stalling. 
%
%
Another method to predict the future bandwidth that has been widely used in the literature is the harmonic mean based prediction \cite{MPC,chen2014msplayer}, which uses the harmonic mean of the past few seconds to predict the bandwidth for the next few seconds.
 
 In this paper, we first theoretically formulate the problem of adaptive-SVC video streaming with the knowledge of future bandwidth. We consider two streaming schemes: {\em skip based} and {\em no-skip based} streaming.
The former is usually for real-time streaming in which there is a playback deadline for each of the chunks, and chunks not received by their respective deadlines are skipped. For no-skip based streaming, if a chunk cannot be downloaded by its deadline, it will not be skipped; instead, a stall (re-buffering) will incur, \ie the video will pause until the chunk is fully downloaded. In both variants, the goal of the proposed scheduling algorithm is to determine up to which layer we need to fetch for each chunk (except for those skipped in realtime streaming), such that the overall quality-of-experience (QoE) is maximized and the number of stalls or skipped chunks is minimized. The key contributions of the paper are described as follows.
\begin{thisnote}

\BULLET A novel metric of QoE is proposed for SVC streaming in both the scenarios (skip and no-skip). The metric is a weighted sum of the layer sizes for each chunk. Since the user's QoE is concave in the playback rate \cite{miller2017qoe}, the higher layers contribute lower to the QoE as compared to the lower layers. Thus, the weights decrease with the layer index  modeling the diminishing returns for higher layers. 



\BULLET We show that even though the proposed problem is a non-convex optimization problem with integer constraints, it can be solved  optimally using an algorithm with a complexity that is linear in the number of chunks.  The proposed algorithm, ``Layered Bin Packing" (LBP) Adaptive Algorithm,  proceeds layer-by-layer, tries to efficiently bin-pack all chunks at a layer and provides maximum bandwidth to the next layer's decisions given the decisions of the lower layers of all the chunks.


\BULLET  We propose an online robust adaptive-SVC streaming algorithm (Online LBP). This algorithm exploits the prediction of the network bandwidth for some time ahead, solves the proposed optimization problem to find the quality decisions for $W$ chunks ahead, and re-runs every $\alpha$ seconds to adjust to prediction errors and find quality decisions for more chunks ahead. 
\end{thisnote}


\BULLET We considered two techniques of bandwidth prediction. First, harmonic mean based prediction which was widely used in the literature \cite{MPC,chen2014msplayer} where the harmonic mean of the past few seconds is used to predict the bandwidth for few seconds ahead (typically 20 seconds ahead). Second, crowd-sourced erroneous bandwidth prediction where bandwidth profiles experienced by people travelled the same road recently are used to predict the bandwidth for the current user.

%

\BULLET Trace-driven simulation using datasets collected from commercial cellular networks demonstrates that our approach is robust to prediction errors, and works well with short prediction windows (\eg 20 seconds). The proposed approach is compared with a number of adaptation strategies including slope based SVC streaming~\cite{SVCMetrics}, Microsoft's smooth streaming algorithm (adapted to streaming SVC content), and Netflix's buffer-based streaming algorithm (BBA-0)~\cite{BBA} (adapted to SVC). 

\BULLET The results demonstrate that our algorithm outperforms the state-of-the-art by improving key quality-of-experience (QoE) metrics such as the playback quality, the number of layer switches, and the number of skips or stalls. 

\BULLET In addition to the simulations, we built a testbed that streams synthetic SVC content over TCP/IP networks using real LTE traces. We then implemented our streaming algorithm on the testbed and evaluated it under challenging network conditions. The emulation outcome is very close to the simulation results and incurs very low run-time overhead,
further confirming that our algorithm can be practically implemented and deployed on today's mobile devices.

\section{Related Work}
\label{sec:related}

Video streaming has received a lot of attention from both the academia and industry in the past decade.
We summarize some of the efforts devoted to streaming technologies that are based on
Adaptive Bit Rate (ABR), Adaptive-SVC, and that rely on network bandwidth prediction.

{\bf{ABR Streaming}.} The recent adoption of the open standards MPEG-DASH ~\cite{DASH} has made ABR streaming the most popular video streaming solution. Commercial systems such as Apple's HLS~\cite{HLS}, Microsoft's Smooth Streaming~\cite{SS}, and Adobe's HDS~\cite{HDS} are all ABR streaming algorithms. In recent studies, researchers have investigated various approaches for making streaming decisions, for example, by using control theory~\cite{MPC,Miller15}, Markov Decision Process~\cite{Jarnikov11}, machine learning~\cite{Claeys14}, client buffer information~\cite{BBA}, and data-driven techniques~\cite{Liu12,C3,CS2P}. In this work, we use an optimization-based approach to design novel streaming algorithms for Adaptive-SVC streaming whose encoding scheme is very different from that of used for ABR streaming.

{\bf{Adaptive-SVC Streaming.}}
SVC encoding received the final approval to be standardized as an amendment of the H.264/MPEG-4  standard
in 2007~\cite{4317636}. 
Although
much less academic research has been conducted on Adaptive SVC streaming compared to ABR streaming, there exist some studies of using SVC encoded videos  to adapt video playback quality to network conditions.
A prior study~\cite{LayeredAdaptation} proposed a server-based quality adaptation mechanism that performs coarse-grained rate adaptation by adding or dropping layers of a video stream.
While this mechanism was designed to be used over UDP with a TCP-friendly rate control, more recent research has explored techniques that use Adaptve-SVC streaming  over HTTP.
A study~\cite{Sanchez12} compared SVC  with regular H.264 encoding (H.264/MPEG).
Their results suggest that SVC outperforms H.264/AVC
for scenarios such as VoD and IPTV through more effective rate adaptation.
The work~\cite{SVCDataset} published the first dataset and toolchain for SVC.
Some prior work~\cite{Andelin12,Sieber13} proposed new rate adaptation algorithms for Adaptive-SVC streaming that prefetch future base layers and backfill current enhancement layers.
Our work differs from the above in that we
develop low-complexity algorithms that explicitly and strategically leverage the future knowledge of network conditions for better rate adaptation. 

{\bf{Streaming that Exploits Network Condition Prediction.}}
The knowledge of the future network conditions can play an important role in Internet video streaming.
%
%
A prior study~\cite{Zou15} investigated the performance gap between state-of-the-art streaming approaches and the approach with accurate bandwidth prediction for ABR.
The results indicate that prediction brings additional performance boosts for ABR, and thus motivates our study.
%
Prior studies~\cite{Riiser12,GTube} proposed ABR streaming mechanisms that use pre-collected geo-tagged network bandwidth profiles.
Our work also exploits the predictable nature of future network conditions, but provides an optimization based framework in the context of SVC-based encoding. In Appendix \ref{sec:predict}, we show more evidence of network predictability in the context of cellular networks.
 We note that even though there is a broad interest in the bitrate adaptation  algorithms, a principled understanding of algorithms is limited. One of the key fundamental approachs to formulate the optimization problem was given in \cite{MPC}. However, the proposed algorithm in \cite{MPC} is computationally hard and thus a lookup table is hard coded based on solving the optimization problem offline for a given set of encoding rates. To make the table size small, the offline solution is divided in coarse bins thus giving an approximate solution. {\cite{CS2P} proposed crowd-source based bandwidth prediction and used the streaming algorithm proposed in~\cite{MPC} to make the bit rate decision per video's chunk. Moreover, \cite{wang2017towards} gives feasible solution by relaxing the integer constraints in the streaming optimization problem.  Further, \cite{zou2015can} considers prediction-based formulation while giving heuristics to solve the problem.}
 
 In contrast to \cite{MPC}, we propose an online algorithm that  solves the optimization problem optimally in linear complexity and can run on the fly. Thus, the proposed approach does not need to hard code information for different encoding rates. Moreover, the offline algorithm is shown to be also optimal and is solvable in linear time complexity. Therefore, we provide a theoretic upper bound to our formulation. Finally, we do not relax any of the constraint, we consider both skip and no-skip based streaming scenarios, and we show optimality in both cases.

\section{System Model}
\label{sec:model}

We consider the problem of adaptively streaming an SVC video. 
An SVC encoded video is divided into $C$ chunks (segments) and stored at a server. Every chunk is of length $L$ seconds, and is encoded in Base Layer (BL) with rate $r_0$ and $N$ enhancement layers ($E_1, \cdots, E_{N}$) with rates $r_1, \cdots, r_{N}$ $\in$  ${\mathcal R} \triangleq \{0,r_0,r_1,\cdots, r_{N}\}$. We assume that each layer is encoded at constant bit rate (CBR). In other words, all chunks have the same $n$th layer size. Let the size of the $n$-th layer of chunk $i$ be $Z_{n,i} \in {\mathcal Z}_n \triangleq \{0,Y_n\}$, where $Y_n=L \times r_n$. Let the size of a chunk that is delivered at the $n$-th layer quality be $X_n(i)$, where $X_n(i)=\sum_{m=0}^n Y_m$.

Let $z_n(i,j)$ be the size of layer $n$ of chunk $i$ that is fetched at time slot $j$, and $x(i,j)$ be what is fetched of all layers of chunk $i$ at time slot $j$, i.e., $x(i,j)=\sum_{n=0}^N z_n(i,j)$.
Further, let $B(j)$ be the available bandwidth at time $j$. For the offline algorithm, we assume the bandwidth can be perfectly predicted. Also let $s$ be the startup delay and $B_m$ be the playback buffer size in time units (\ie the playout buffer can hold up to $B_m$ seconds of video content). We assume all time units are discrete and the discretization time unit is assumed to be 1 second (which can be scaled based on the time granularity). Since the chunk size is $L$ seconds, the buffer occupancy  increases by $L$ seconds when chunk $i$ starts downloading (we reserve the buffer as soon as the chunk start downloading).

{The optimization framework can run at either the client or the server side as long as the required inputs are available. {A setup where the algorithm is run at the client side is depicted in Fig.~\ref{fig:encoding}.}  The algorithm takes as an input, the predicted bandwidth for the time corresponding to the next $C$ chunks, layer sizes ($Y_0,...,Y_N$), startup delay ($s$), and maximum buffer size $B_m$, and outputs the layers that can be requested for the next $C$ chunks ($Z_{n,i},  i \in \{1,..C\}, n \in \{0,...,N\}$). The video chunks will be fetched according to the requested policy and in order. For the online algorithm, this process repeats every $\alpha$ seconds, and decisions can be changed on fly since the proposed algorithm  adapts to the prediction error.}

\begin{figure}[ht]
	\centering
	\small
	\includegraphics[trim=0in 1.6in 0in 1.2in, clip,width=.48\textwidth]{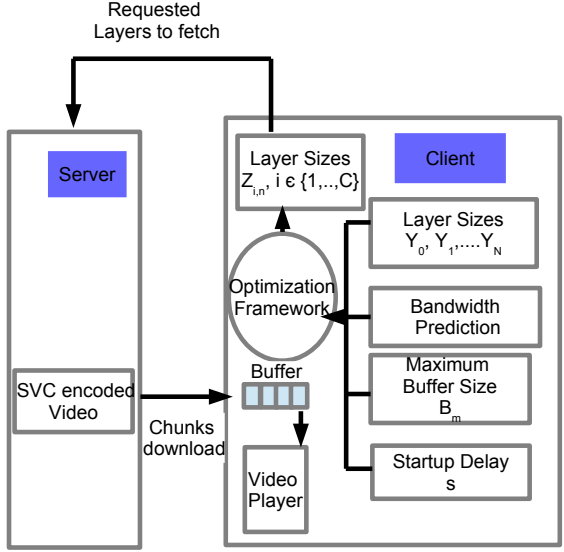}
	\vspace{-.3in}
	\caption{System Model}
	\label{fig:encoding}
	\vspace{-.1in}
\end{figure}

We consider two scenarios: skip based streaming and no-skip based streaming. For skip streaming, the video is played with an initial start-up (\ie buffering) delay $s$ seconds  and there is a playback deadline for each of the chunks where chunk $i$ need to be downloaded by time $deadline(i)$. Chunks not received by their respective deadlines are skipped. For no-skip streaming, it also has start-up delay. However, if a chunk cannot be downloaded by its deadline, it will not be skipped. Instead, a stall (\ie rebuffering) will occur \ie the video will pause until the chunk is fully downloaded.
In both scenarios, the goal of the scheduling algorithm to be detailed next is to determine up to which layer we need to fetch for each chunk (except for those skipped), such that  the number of stalls or skipped chunks is minimized as the first priority, the overall playback bitrate is maximized as the next priority, the number of quality switching between neighboring chunks is minimized as the third priority. Similar to many other studies on DASH video streaming \cite{MPC, zou2015can, BBA}, 
this paper does not consider mean opinion score (MOS) metric since obtaining MOS ratings are video-dependent and are time-consuming and expensive as they require recruitment of human assessors. A table of notations used in this paper is included in Appendix~\ref{sec:notationTable}.
\section{Adaptive SVC Streaming}
\label{sec:alg}

We now detail the adaptive SVC streaming algorithms. We describe the basic formulation for skip-based streaming in~\S\ref{sec:problem}. We then identify the particular problem structure in our formulation and strategically leverage that to design a linear-time solution in~\S\ref{sec:opt} and~\S\ref{skipalgo}. 
We prove the optimality of our solution in~\S\ref{sec:proof}. An example of the algorithm is given in Appendix \ref{skipex}, and detailed proofs are in Appendix \ref{apdx:skip:beta1}.  We then extend the basic scheme to its online version in~\S\ref{sec:bw_err} and to no-skip based streaming in~\S\ref{sec:noskip} (with detailed algorithm in Appendix \ref{sec: noSkipCode}, example in Appendix \ref{noskipex}, and proofs in Appendix \ref{apdx:lemma1Noskip}).



\subsection{Skip Based Streaming: Offline Problem Formulation}
\label{sec:problem}

Given the settings described in~\S\ref{sec:model}, we first formulate an offline optimization problem. It
jointly \emph{(i)} minimizes the number of skipped chunks, \emph{(ii)} maximizes the average playback rate of the video, and \emph{(iii)} minimizes the quality changes between the neighboring chunks to ensure the perceived quality is smooth.
We give a higher priority to \emph{(i)} as compared to \emph{(ii)}, since skips cause more quality-of-experience (QoE) degradation compared to playing back at a lower quality~\cite{MPC}. Further,  \emph{(iii)} is the lowest priority among the three objectives.  {The proposed formulation maximizes a weighted sum of the layer sizes. The weights are along two directions. The first is across time where the layers of the later chunks are weighed higher using a factor $\beta\ge 1$. The second is across the layers where fetching the $n$-th layer of a chunk achieves a utility that is $0<\gamma<1$ times the utility that is achieved by fetching the $(n-1)$-th layer. Thus, the objective is given as $\sum_{n=0}^{N}\gamma^n\sum_{i=1}^{C}\beta^i Z_{n,i}$}. We further assume that

\begin{equation}
\gamma^a r_a > \sum_{k=a+1}^N\gamma^kr_k \sum_{i=1}^C \beta^i \  \text{  for } a= 0, \cdots, N-1. \label{basic_gamma_1}
\end{equation}
This choice of $\gamma$ implies that all the higher layers than layer $a$ have lower utility than a chunk at layer $a$ for all $a$. For $a=0$, this implies that all the enhancement layers have less utility than a chunk at the base layer. Thus, the avoidance of skips is the highest priority.
The use of $\gamma$ helps prioritize lower layers over higher layers and models concavity of user QoE with playback rate. Due to this weight, the proposed algorithm will avoid skip as the first priority and will not use the bandwidth to fetch higher layers at the expense of base layer. Similar happens at the higher layers. The combination of the two weights help minimize multi-layer quality switches between neighboring chunks since the use of $\gamma$ discourages getting higher layers at the expense of lower layers. We assume $\beta=1+\epsilon$ where $\epsilon > 0$ is very small number (\eg 0.001). {  The use of $\beta= 1+\epsilon$ helps in three aspects, (i) {makes} optimal layer decisions for different chunks unique, (ii) better adaptability to the bandwidth fluctuations by {preferring fetching higher layers of later chunks}, and (iii) reduction of quality variations.}  Indeed, if the playback buffer is not limited, there will ideally be a few jumps of quality increases and no quality decrease in the playback of the chunks using this metric. {  An example to further explain the objective and the above mentioned points for $\gamma$ and $\beta$ is provided in Appendix~\ref{sec:formulationExample}.}

Overall, the SVC layer scheduling problem with the knowledge of future bandwidth information
can be formulated as follows, where ${\bf I}(.)$ is an indicator function which has the value $1$ if inside expressions holds and zero otherwise.
\begin{eqnarray}
\textbf{Maximize: }\Bigg(\sum_{n=0}^{N}\gamma^n\sum_{i=1}^{C}\beta^i Z_{n,i}\Bigg)
\label{equ:eq1}
\end{eqnarray}
subject to
\begin{eqnarray}
\sum_{j=1}^{(i-1)L+s} z_n(i,j) = Z_{n,i},\quad  \forall i,  n
\label{equ:c2eq1}
\end{eqnarray}
\begin{eqnarray}
 Z_{n,i}\le \frac{Y_n}{Y_{n-1}}Z_{n-1,i},\quad  \forall i,  n
\label{equ:c2eq11}
\end{eqnarray}
\begin{eqnarray}
\sum_{n=0}^N\sum_{i=1}^{C} z_n(i,j)  \leq B(j) \  \   \forall j=1, \cdots, (C-1)L+s,
\label{equ:c3eq1}
\end{eqnarray}
\begin{eqnarray}
\sum_{i, (i-1)L+s > t} {\bf I}\Bigg(\sum_{j=1}^t\bigg(\sum_{n=0}^Nz_n(i,j)\bigg)> 0\Bigg) L \leq B_m \   \forall t
\label{equ:c4eq1}
\end{eqnarray}
\begin{equation}
z_n(i,j) \geq 0\   \forall i = 1, \cdots, C
\label{equ:c5eq1}
\end{equation}
\begin{equation}
z_n(i,j) = 0\   \forall i, j > (i-1)L+s
\label{equ:c6eq1}
\end{equation}
\begin{equation}
Z_{n,i} \in {\mathcal Z}_n \quad  \forall i = 1, \cdots, C, \text{ and } \forall n = 1, \cdots, N
\label{equ:c7eq1}
\end{equation}

\begin{eqnarray}
\text{Variables:}&& z_n(i,j), Z_{n,i} \ \ \  \forall   i = 1, \cdots, C,  \nonumber \\
&&j = 1, \cdots, (C-1)L+s, \  n = 0, \cdots, N \nonumber
\end{eqnarray}

%
Constraints \eqref{equ:c2eq1} and \eqref{equ:c7eq1} ensure that what is fetched for any layer $n$ of a chunk $i$ over all times to be either zero or the $n$-th layer size. The decoder constraint  \eqref{equ:c2eq11} enforces that the $n$th layer of a chunk cannot be fetched if the lower layer is not fetched since this layer will not be decoded because of the layer dependency.
\eqref{equ:c3eq1} imposes the available bandwidth constraint at each time slot $j$ and \eqref{equ:c4eq1} imposes the playback buffer constraint so that the content in the buffer at any time does not exceed the buffer capacity $B_m$.
Constraint \eqref{equ:c5eq1} imposes the non-negativity of the chunk download sizes, and \eqref{equ:c6eq1} enforces not to fetch a chunk after its deadline. The deadline of chunk $i\in \{1, \cdots, C\}$ is  $deadline(i)=(i-1)L+s$.

\subsection{Optimization Problem Structure}
\label{sec:opt}

The problem defined in~\S\ref{sec:problem} has integer constraints and has an indicator function in a constraint. This problem is in the class of combinatorial optimization \cite{nemhauser1988integer}. Some of the problems in this class are the Knapsack problem, Cutting stock problem, Bin packing problem, and Travelling salesman problem. These problems are all known to be NP hard. Very limited problems in this class of combinatorial optimization are known to be solvable in polynomial time. Some typical examples being  shortest path trees, flows and circulations, spanning trees, matching, and matroid problems. The well known  Knapsack problem optimizes a linear function with a single linear constraint ( for integer variables), and is known to be NP hard. The optimization problem defined in this paper has multiple constraints, and does not lie in any class of known combinatorial problems that are polynomially-time solvable to the best of our knowledge. In this paper, we will show that this combinatorial optimization problem can be solved optimally in polynomial time. 


\subsection{Optimal Linear-time Solution}\label{skipalgo}

We now show the proposed problem in ~(\ref{equ:eq1}-\ref{equ:c7eq1}) can be solved optimally with a complexity of $O(CN)$. 
%
We call our proposed algorithm ``Layered Bin Packing Adaptive Algorithm" (LBP), which is summarized in Algorithm 1. At a high level,
our algorithm works from the lowest (\ie the base) to the highest enhancement layer, and processes each layer separately. It performs backward and forward scans (explained below) at each layer given the decisions of the previous layers.


Running the backward scan at the $n$th layer (Algorithm 2) finds the maximum number of chunks that can be fetched up to the $n$th layer quality given the decisions of the previous layers. Then, running the forward  scan (Algorithm 3) simulates fetching chunks in sequence  as early as possible, so the start time of downloading chunk $i$  (the lower deadline $t(i)$) is found. Lower and Upper ($t(i)$, $deadline(i)$) deadlines will be used to find the next layer decisions (as explained below). 
%
%
\anis{now?. I removed many lines, since backward and then forward algo are explained in details bellow}\feng{Check the above. Give some intuitions of why this approach works - that is critical.} 



{\bf Backward Algorithm for Base Layer: } Given the bandwidth prediction, chunk deadlines, and the buffer size, the algorithm simulates fetching the chunks at base layer quality starting from the last towards the first chunk. The deadline of the last chunk is the starting time slot of the backward algorithm scan. \anis{now?}\feng{starting time of what? unclear.}The goal is to have chunks fetched closer to their deadlines. \anis{now?}\feng{``close'' -- you mean ``right before?''}For every chunk $i$, the backward algorithm checks the bandwidth and the buffer; if there is enough bandwidth and the buffer is not full, then chunk $i$ is selected to be fetched (line 18-22). 
\feng{``candidate'' and ``can be fetched'' are different. Do you mean ``it is selected to be fetched''?}
The algorithm keeps checking this feasibility to select chunks to be fetched.  
If a chunk $i^\prime$ is not selected to be fetched, one of the following two scenarios could have happened. The first scenario is the violation of the {\bf buffer} capacity, where selecting the chunk to be fetched would violate the playback buffer constraint.  The second is the {\bf bandwidth} constraint violation where the remaining available bandwidth 
is not enough for fetching a chunk. This scenario also means that the chunk could not be fetched by its deadline, so it can also be called deadline violation.

For buffer capacity violation, we first note that, there could be a chunk  $i^{\prime\prime} > i^\prime$ in which if it is skipped, chunk  $i^{\prime}$ can still be fetched. However, the backward algorithm decides to skip downloading chunk $i^\prime$ (line 8). We note that since there is a buffer capacity violation, one of the chunks must be skipped. The reason of choosing to skip chunk $i^\prime$ rather than a one with higher index is that $i^\prime$ is the closest to its deadline. Therefore, $i^\prime$ is not better candidate to the next layer than any of the later ones.\anis{now?}
\feng{confusing - you said ``skipping downloading the chunk'', then ``choosing the current chunk''?}
In the second case of deadline/bandwidth violation, the backward algorithm decides to skip chunks up to $i^\prime$ \feng{which chunk?}since there is not enough bandwidth. As before, since equal number of chunks need to be skipped anyway, skipping the earlier ones is better because it helps in increasing the potential of getting higher layers of the later chunks\anis{now?}. \feng{the last sentence is unclear and doesn't parse.}



{\bf Forward Algorithm for Base Layer: } The forward algorithm takes the chunk size decisions from the Backward step which provides the base layer size decision of every chunk $i$ which is either 0 or the BL size. Then, the forward algorithm simulates fetching the chunks in sequence starting from the first one. Chunks are fetched as early as possible with the deadline, buffer, and the bandwidth constraints being considered. 
The chunks that were not decided to be fetched by the Backward Algorithm are skipped (any chunk $i \notin I_0$, line 6 ). The forward algorithm provides the the earliest time slot \feng{the earliest time?}when chunk $i$ can be fetched ($t(i)$, line 10). This time is used as a {\bf lower deadline} on the time allowed to fetch chunk $i$ when the backward algorithm is run for the next layer. Therefore, the backward size decisions of base layer of earlier chunks can not be violated when the backward algorithm is re-run for deciding the first enhancement layer sizes ($E1$ decisions). \anis{now?}\feng{unclear what ``lower bound'' means. The later sentence is also unclear.}
Moreover, it provides the portion \feng{the amount of what?} that can be fetched of chunk $i$ at its lower deadline $t(i)$ ($a(i)$, line 11) and the remaining  bandwidth at every time slot $j$ after all non skipped chunk are fetched ($e(j)$, line 12). 
\anis{now?}\feng{Overall this paragraph is confusing and unclear. Needs revision.}


 {\bf Modifications for Higher Layers: } The same backward and forward steps are used for each layer given the backward-forward decisions of the previous one on the chunk sizes and lower deadlines. The key difference when the algorithm is run for the enhancement layer decisions as compared to that for the base layer is that the higher layer of the chunk is skipped if the previous layer is not decided to be fetched. When running the backward algorithm for $E1$ decisions, for every chunk $i$, we consider the bandwidth starting from the lower deadline of that chunk $t(i)$, so previous layer decisions (base layer decisions) of early chunks can't be violated. The same procedure is used to give higher layer decisions when all of the lower layer decisions have already been made.  An  example to illustrate the algorithm is given in Appendix \ref{noskipex}. 
\anis{removed}\feng{the last sentence is not clear.}

\begin{figure}[htbp]
		\begin{minipage}{\linewidth}
			\begin{algorithm}[H]
				\small
				\begin{algorithmic}[1]
				\STATE {\bf Input:} $Y_n$, $deadline(i)$, $s$, $B_m$, $C$, $B(j)$: available bandwidth at time $j$,
									\STATE {\bf Output:} $X(i) \forall i$: The maximum size in which chunk $i$ can be fetched, $I_n$: set contains the indices of the chunks that can be fetched up to layer $n$ quality.
					
    \STATE {\bf Initialization:}
   \STATE $X_n=\sum_{m=0}^n Y_m$ cumulative size up to  layer $n$
    \STATE $c(j)=\sum_{j^\prime=1}^{j} B(j^\prime)$ cumulative bandwidth up to time $j, \forall j$
   \STATE $t(i)=0, \forall i$, first time slot chunk $i$ can be fetched
   \STATE $a(i)=0, \forall i$, lower layer decision of fetched amount of chunk $i$ at its lower deadline time $t(i)$
   \STATE $e(j)=B(j), \forall j$,  remaining  bandwidth at time j after all non skipped chunk are fetched according to lower layer size decisions 
   \STATE $X(i)=0, deadline(i)=(i-1)L+s \quad \forall i$ 
   \STATE $bf(j)=0, \forall j$, buffer length at time $j$
    \STATE {\bf For each  layer, $n = 0, \cdots, N$}
      \STATE $[X, I_n] = backwardAlgo(B, X,X_n, C, L, deadline, B_m,bf,$   $t,c,a,e)$
     \STATE $[t,a,e] = forwardAlgo(B, X, C, deadline, B_m, bf,I_n)$


   				\end{algorithmic}
				\caption{Layered Bin Packing Adaptive Algorithm }
			\end{algorithm}
		\end{minipage}
	\end{figure}

\begin{figure}[htbp]
		\begin{minipage}{\linewidth}
			\begin{algorithm}[H]
				\small
				\begin{algorithmic}[1]
				\STATE {\bf Input:} $B, X, X_n, C, L, deadline, B_m,bf,t,c,a,e$
					\STATE {\bf Output:} $X(i)$ size of chunk $i$, $I_n$: set contains chunks that can be fetched in quality up to $n^{th}$ layer.
    \STATE {\bf Initilization:}
    \STATE $i=C$, $ j=deadline(C)$
   \STATE initialize $bf(j)$ to zeros $\forall j$.

     \WHILE {($j > 0$ and $i > 0$)}
        \IF{$j <= deadline(i)$}
        \STATE {\bf if} {$(bf(deadline(i)) = B_m)$} {\bf then } $i=i-1$

                \IF{$j$ is the first time to fetch chunk $i$ from back}
                		\IF{$(t(i)=0)$}
				\STATE $rem1=c(j)-c(1)+e(1)$, $rem2=rem1$
			\ELSE
			\STATE $rem2=c(j)-c(t(i))$, $rem1=rem2+e(t(i))+a(i)$
				
			\ENDIF
                		\IF{$(rem1 < X_n(i))$}
				\STATE {\bf if} {$(X(i) > 0)$} {\bf then} $X_n(i)=X(i)$ {\bf else } $i=i-1$
			 \ELSE
		        		
                			\IF{$(rem2 < X_n(i))$ and $rem1 \geq X_n(i))$}
                				\STATE $e(t(i))=e(t(i))+rem1-X_n$
				\ENDIF
			        \STATE $X(i)=X_n(i)$, $I_n \leftarrow I_n\cup i$
            		\ENDIF
          	\ENDIF
	
            \STATE $fetched=min(B(j), X_n(i))$, $B(j)=B(j)-fetched$ 
		\STATE $X_n(i)=X_n(i)-fetched$
            \STATE  {\bf if} {$(X_n(i) > 0)$} {\bf then } $bf(j)=bf(j)+L$
            	
            \STATE {\bf if } {$(X_n(i)=0)$} {\bf then } $i=i-1$
	
	   \STATE {\bf if } {$(B(j)=0)$} {\bf then }  $j=j-1$            	
            \ELSE
           \STATE $j=j-1$
        \ENDIF
\ENDWHILE
				\end{algorithmic}
				\caption{Backward Algorithm }
			\end{algorithm}
		\end{minipage}
		\vspace{-.2in}
	\end{figure}

\begin{figure}[htbp]
		\begin{minipage}{\linewidth}
			\begin{algorithm}[H]
				\small
				\begin{algorithmic}[1]
				\STATE {\bf Input:} $B, X, C, deadline, Bm, bf,I_n$

				\STATE {\bf Output:} $t(i)$: first time slot chunk $i$ can be fetched (lower deadline of chunk $i$), $a(i)$, decision of fetched amount of chunk $i$ at its lower deadline time slot $t(i)$, $e(j)$,  remaining  bandwidth at time j after all non skipped chunk are fetched according to the decided layer size. 

   \STATE $ j=1$, $k=1$
    \WHILE {$j \leq deadline(C)$ and $k \leq max(I_0)$ (last chunk to fetch)}
	\STATE $i=I(k)$
          \STATE {\bf if }{$i=0$} {\bf then } $k=k+1$
            \IF{$j \leq deadline(i)$}
             \STATE {\bf if } {$(bf(j) = B_m)$} {\bf then }  $j=j+1$

                 \STATE $fetched=min(B(j), X(i))$

                \IF{$j$ is the first time chunk $i$ is fetched}
                 \STATE $t(i)=j$,
                 \STATE $a(i)=fetched$ 
                    \ENDIF
                 \STATE $B(j)=B(j)-fetched$
                 \STATE $e(j)=B(j)$, $X(i)=X(i)-fetched$

                   \STATE {\bf if } {$X(i) > 0$} {\bf then }  $bf(j)=bf(j)+L$
                  \STATE {\bf if } {$X(i)=0$} {\bf then } $k=k+1$

                   \STATE {\bf if } {$B(j)=0$} {\bf then } $j=j+1$
               \ELSE
                \STATE $k=k+1$
            \ENDIF

   \ENDWHILE

				\end{algorithmic}
				\caption{Forward Algorithm }
			\end{algorithm}
		\end{minipage}
		\vspace{-.2in}
	\end{figure}



{\bf Complexity Analysis}: The initialization clearly sums the variables over time, and is at most O($C$) complexity.  At each layer, a backward and a forward algorithm are performed. Both the algorithms have a while loop, and within that, each step is O(1). Thus, the complexity is dependent on the number of times this loop happens. For the backward algorithm, each loop decreases either $i$ or $j$ and thus the number of times the while loop runs is at most $C+deadline(C)+1$. Similarly, the forward algorithm while loop runs at most  $C+deadline(C)+1$ times. In order to decrease the complexities,  cumulative bandwidth for every time slot $t$, $r(t)$ is used to avoid summing over the bandwidth in the backward and the forward loops.


{\bf Adaptation to ABR Streaming}: We note that the proposed algorithm selects quality levels for every chunk and can also be used for ABR streaming. For a given set of available ABR
rates, the difference between the rates for the coded chunk at quality level $n+1$ and quality level n can
be treated as the nth layer SVC rate for all $n$.



\subsection{Optimality of the Proposed Algorithm} 
\label{sec:proof}
In this subsection, we prove the optimality of Layered Bin-Packing Adaptive Algorithm in solving the optimization problem ~(\ref{equ:eq1}-\ref{equ:c7eq1}). We first note that it is enough to prove that the algorithm is the best among any in-order scheduling algorithm (that fetches chunks in order based on the deadlines). This is because for any other feasible fetching algorithm, we can convert it to an in-order fetching algorithm with the same bandwidth utilizations for each chunk. Getting in-order  helps the buffer and other constraints. Thus, we can obtain the same objective and can satisfy the constraints.  The following Lemma  states that given the lower and upper deadlines (($t(i)$) and $deadline(i)$) of every chunk $i$, the $(n-1)th$ layer quality decision, running the backward algorithm for the $n$th layer maximizes the number of chunks that can have their $n$th layer fetched.

\begin{lemma}
Given size decisions up to ($n-1$)th layer, and lower and upper deadlines ($t(i)$, and $deadline(i)$) for every chunk $i$, the backward algorithm achieves the minimum number of the $n$th layer skips as compared to any feasible algorithm which fetches the same layers to every chunk up to the layer $n-1$.
 \label{lem:skip:beta1}
\end{lemma}
\begin{proof}
Proof is provided in  Appendix.~\ref{apdx:skip:beta1}.
\end{proof}
The above lemma shows that backward algorithm minimizes the $n$th layer skips given the lower and upper deadlines of every chunk. However, it does not tell us if that lower deadline is optimal or not. The following proposition shows that for any quality decisions, the forward algorithm finds the optimal lower deadline on the fetching time of any chunk. 


\begin{proposition}
if $t_f(i)$ is the earliest time to start fetching chunk $i$ using the forward algorithm (lower deadline), and $t_x(i)$ is the earliest time to fetch it using any other in sequence fetching algorithm, then the following holds true.
$$t_f(i) \leq t_x(i).$$
\label{noSkipPro}
\end{proposition}

The above proposition states that $t_f(i)$ is the lower deadline of chunk $i$, so chunk $i$ can't be fetched earlier without violating size decisions of the lower layers of earlier chunks. Therefore, at any layer $n$, we are allowed to increase the chunk size of chunk $i$ as far as we can fully fetch it within the period between its lower and upper deadlines. If increasing its size to the $n$-th layer quality level requires us to start fetching it before its lower deadline, then we should not consider fetching the $n$-th layer of this chunk . Fetching the $n$-th layer of this chunk in this case will affect the lower layer decisions and will cause dropping lower layers of some earlier chunks. Since, our objective prioritizes lower layers over higher layers ($0<\gamma < 1$ and \eqref{basic_gamma_1}), lower deadline must not be violated. As a simple extension of Lemma \ref{lem:skip:beta1}, we can consider any $\beta\ge 1$. 

\begin{lemma}
Given optimal solution of layer sizes up to the ($n-1$)th layer, and lower and upper deadlines ($t(i)$, and $deadline(i)$) of every chunk $i$. If ${\mathcal  Z}_n^* = (Z_{n,i}^* \forall n, i )$ is the $n$-th layer solution that is found by running the backward algorithm for the $n$th layer for the $n$th layer sizes, and ${\mathcal  Z}^\prime_n = (Z_{n,i}^\prime \forall n, i )$ is a feasible solution that is found by running any other algorithm, then the following holds for any $\beta \geq 1$. 
\begin{equation}
\sum_{i=1}^{C}\beta^i Z_{n,i}^\prime \leq \sum_{i=1}^{C}\beta^i Z_{n,i}^*
\label{equ:eq1_lemma1}
\end{equation}
\label{lem:skip:betage1}
\end{lemma}
\begin{proof}
Proof is provided in the Appendix.~\ref{apdx:skip:betage1}. 

\end{proof}

We note that Lemma \ref{lem:skip:beta1} is a corollary of Lemma \ref{lem:skip:betage1}, which can be obtained when $\beta=1$.

Using Lemma.~\ref{lem:skip:beta1}, Proposition.~\ref{noSkipPro}, and Lemma.~\ref{lem:skip:betage1}, we are ready to show the optimality of Layered Bin Packing Adaptive Algorithm in solving problem~(\ref{equ:eq1}-\ref{equ:c7eq1}), and this is stated in the following theorem.

\begin{theorem}
Up to a given enhancement layer $M, M \geq 0$, if $Z_{m,i}^*$ is the size of every layer $m \leq M$ of chunk $i$ that is found by running Layered Bin Packing Adaptive Algorithm, and $Z_{m,i}^\prime$ is the size that is found by running any other feasible algorithm, then the following holds for any $0<\gamma < 1$, satisfies (\ref{basic_gamma_1}), and $\beta \le 1$.
\begin{equation}
 \sum_{m=0}^M \gamma^m\sum_{i=1}^{C}\beta^i Z_{m,i}^\prime \leq \sum_{m=0}^{M} \gamma^m\sum_{i=1}^{C}\beta^i Z_{m,i}^*.
\label{equ:thm}
\end{equation}
In other words, Layered Bin Packing Adaptive Algorithm achieves the optimal solution of the optimization problem~(\ref{equ:eq1}-\ref{equ:c7eq1}) when $0<\gamma < 1$,  satisfy (\ref{basic_gamma_1}), and $\beta \ge 1$.
\label{theorem: theorem1}
\end{theorem}
\begin{proof}
Proof is provided in the Appendix~\ref{apdx:them1}.
\end{proof}

\subsection{Online Algorithm: Dealing with Short and Inaccurate BW Prediction}
\label{sec:bw_err}

We face two issues in reality. First, the bandwidth information for the distant future may not always be available. 
Second, even for the near future, the estimated bandwidth may have errors. 
To address  both of these challenges, we design an online algorithm (Algorithm \ref{onlineskip}). The algorithm works as follows.
Every $\alpha$ seconds, we predict the bandwidth for $W$ seconds ahead (lines 9-10). Typically $\alpha$ is much smaller than $W$ ($\alpha \ll W$). 
 We find the last chunk to consider in this run of the algorithm (line 11). The online algorithm thus computes the scheduling decision only for the chunks corresponding to the next $W$ seconds ahead. We  re-compute the quality decisions periodically (every $\alpha$ seconds) in order to adjust to any changes in the prediction. We can also run the computation after the download of every chunk (or layer) due to the low complexity of our algorithm.  
Moreover, to handle inaccurate bandwidth estimation, we set lower buffer threshold $(B_{min})$, so if the buffer is running lower than this threshold, we reduce the layer decision by 1 (except if a chunk is already at base layer quality) (lines 15-16). In the real chunk download, if we are within a certain threshold from the deadline of the current chunk and it is not yet fully downloaded, we stop fetching the remaining of the chunk as far as the base layer is fetched and we play it at the quality fetched so far. 

\begin{figure}
		\begin{minipage}{\linewidth}
			\begin{algorithm}[H]
				\small
				\begin{algorithmic}[1]
				\STATE {\bf Input:}  $Y_n$, $deadline(i)$, $s$, $B_m$, $C$, $B(j)$, $W$: the prediction window size, $\alpha$: the decision reconsideration period. 
									\STATE {\bf Output:} $X(i) \forall i$: The maximum size in which chunk $i$ can be fetched, $I_n$: set contains the indices of the chunks that can be fetched up to layer $n$ quality.
					
    \STATE {\bf Initialization:}
   \STATE same as Algorithm 1, offline version plus the following:
  \STATE $sc=1$, the index of the chunk to start with.
  \STATE $ec=1$, the index of the last chunk to consider.
   \STATE $st=1$, the current time slot.
   \STATE {\bf Every $\alpha$ seconds do:}
   \STATE $\quad$ collect user position and speed.
   \STATE $\quad$ predict the bandwidth for $W$ seconds ahead.
   \STATE $\quad$ $ec=$The index of the first chunk has its deadline $\geq st+W$ 
    \STATE $\quad$ {\bf For each  layer, $n = 0, \cdots, N$}
      		\STATE $\quad$$\quad$ $[X, I_n] = backwardAlgo(B, X,X_n, sc, ec, L, deadline,$ $B_m,bf,t,c,a,e)$
     		\STATE $\quad$$\quad$ $[t,a,e] = forwardAlgo(B, X, sc, ec, deadline, Bm,$  $ bf,I_n)$
     	\STATE $\quad$ $sc=$last fetched chunk$+1$
	\STATE $\quad$ $st=$current time slot
   				\end{algorithmic}
				\caption{Online Layered Bin Packing Adaptive Algorithm }\label{onlineskip}
			\end{algorithm}
		\end{minipage}
		\vspace{-.2in}
	\end{figure}

\subsection{No-Skip Based  Streaming Algorithm}
\label{sec:noskip}



In No-Skip streaming (\ie watching a pre-recorded video), when the deadline of a chunk cannot be met,
rather than skipping it, the player will stall the video and continue downloading the chunk.
%
The objective here is to maximize the weighted sum of the layer sizes while minimizing the stall duration (the rebuffering time). The objective function is slightly different from equation.~\eqref{equ:eq1} since we do not allow to skip the base layers. However, we still allow for skipping the higher layers. 
For the constraints, all constraints are the same as skip based optimization problem except that we introduce constraint~\eqref{equ:c1eq2} to enforce the $Z_0(i)$ for every chunk $i$ to be equal to the BL size ($Y_0$). 
We define the total stall (re-buffering) duration from the start till the play-time of chunk $i$ as $d(i)$. Therefore, the deadline of any chunk $i$ is $(i-1)L+s+d(i)$. The No-Skip formulation can thus be written as:
\begin{eqnarray}
\textbf{Maximize: } \sum_{n=1}^{N}\gamma^n\sum_{i=1}^{C}\beta^i Z_{n,i}-\lambda d(C)
\label{equ:eq2}
\end{eqnarray}
subject to,
\begin{eqnarray}
\sum_{j=1}^{(i-1)L+s+d(i)} z_0(i,j) = Y_{0}   \forall i = 1, \cdots, C
\label{equ:c1eq2}
\end{eqnarray}
\begin{eqnarray}
\sum_{j=1}^{(i-1)L+s+d(i)} z_n(i,j) = Z_{n,i}, \quad  \forall i,  n>0
\label{equ:c2eq2}
\end{eqnarray}
\begin{eqnarray}
 Z_{n,i} \le \frac{Y_n}{Y_{n-1}}Z_{n-1,i}, \quad  \forall i,  n
\label{equ:c2eq21}
\end{eqnarray}
\begin{eqnarray}
\sum_{n=0}^N\sum_{i=1}^{C} z_n(i,j)  \leq B(j) \  \   \forall 1\le j\le (C-1)L+s+d(C),
\label{equ:c3eq2}
\end{eqnarray}
\begin{eqnarray}
\sum_{n=0}^N\sum_{i, (i-1)L+s+d(i) > t}{\bf I}\Bigg(\sum_{j=1}^t\bigg(z_n(i,j)\bigg)> 0\Bigg) L \leq B_m \   \forall t
\label{equ:c4eq2}
\end{eqnarray}
\begin{equation}
z_n(i,j) \geq 0\   \forall i = 1, \cdots, C
\label{equ:c5eq2}
\end{equation}
\begin{equation}
z_n(i,j) = 0\   \forall i, j > (i-1)L+s+d(i)
\label{equ:c6eq2}
\end{equation}

\begin{equation}
d(i+1) \geq d(i)\geq 0\   \forall i = 1, \cdots, C-1 \label{deq}
\end{equation}

\begin{equation}
Z_{n,i} \in {\mathcal Z}_n \quad  \forall i = 1, \cdots, C, and\ \forall n = 1, \cdots, N
\label{equ:c7eq2}
\end{equation}

\begin{eqnarray}
\text{Variables:}&& z_n(i,j), Z_{n,i}, d(i) \forall   i = 1, \cdots, C,  \nonumber \\
&& 1\le j \le (C-1)L+s+d(C), n = 0, \cdots, N \nonumber
\end{eqnarray}

{ This formulation converts multi-objective optimization problem with the
stall duration and weighted quality as the two parameters into a single objective using a tradeoff parameter $\lambda$.}  
$\lambda$ is chosen such that avoidance of one stall is preferred as compared to fetching all the layers of all chunks since users tend to care more about not running into rebuffering over better quality. Specifically, $\lambda$  satisfies the following equation.

\begin{equation}
\lambda >  \sum_{n=0}^N \gamma^n Y_n\sum_{i=1}^C\beta^i
\label{lambda_cond}
\end{equation}

With this assumption, we can solve the optimization problem optimally with a slight modification to the algorithm proposed for the skip based streaming version. The proposed algorithm for the No-Skip version is referred to by  ``{\em No-Skip Layered Bin Packing Adaptive Algorithm}" (No-Skip LBP, Algorithm 5 in Appendix \ref{sec: noSkipCode}). There are a few key differences in this algorithm as compared to the skip version, and we explained them below.

One difference as compared to the skip version is that the first step is to determine the minimum stall time since that is the first priority.  In order to do this, we simulate fetching chunks in order at BL quality (Base layer forward algorithm, Algorithm 6 in  Appendix \ref{sec: noSkipCode}). We first let  $d(1)=\cdots = d(C)=0$. We start to fetch chunks in order. If chunk $i$ can be fetched within its deadline $((i-1)L+s+d(i))$, we move to the next chunk (line 20-21). If chunk $i$ cannot be fetched by its deadline, we continue fetching it till it is completely fetched, and the additional time spent in fetching this chunk is added to $d(k)$ for every $k\ge i$ since there has to be an additional stall in order to fetch these chunks (line 22-24). Using this, we obtain the total stall and the deadline of the last chunk $(d(C)$, and $deadline(C))$ The stall duration of the last chunk (chunk C) gives the total stall duration for the algorithm.

The other difference is in running the backward algorithm for the base layer decisions (see base layer backward algorithm, Algorithm 7 in  Appendix \ref{sec: noSkipCode}). The key difference in running the backward algorithm for the base layer with compare to the skip version is that there must be no BL skips. With the backward algorithm, we will work on moving stalls as early as possible. We run the base layer backward algorithm starting at time slot $j=deadline(C)= (C-1)L+s+d(C)$. The scenario of deadline violation cannot happen due to the procedure of forward step before this. Thus, the possibility of buffer constraint violation must be managed. If we reach a chunk in which there is a buffer constraint violation, we decrement its deadline by 1 and check if the violations can be removed. This decrement can be continued until the buffer constraint violation is avoided (lines 11, 28-29). This provides the deadlines of the different chunks such that stall duration is at its minimum and stalls are brought to the earliest possible time, so we get minimum number of stalls and optimal stall pattern. When stalls are brought to their earliest possible, all chunks can have more time to get their higher layers without violating any of the constraints. Therefore, we have higher chance of getting higher layers of later chunks. Forward algorithm (Algorithm 3) is run after that to simulate fetching chunks in order and provide lower deadlines of chunks for the E1 backward run. For enhancement layer decisions, the backward-forward scan is run as in the skip version case since skips are allowed for the enhancement layers. The main algorithm that calls the forward and backward scans in the sequence we described is ``{\em No-Skip Layered Bin Packing Adaptive Algorithm}" (Algorithm 5). An illustrative example of the algorithm is described in Appendix \ref{noskipex}.

\begin{lemma}
If $d^*(C)$ is the total stall duration that is found by No-Skip base layer forward algorithm and $d^{\prime}(C)$ is the total stall duration that is found by running any other feasible algorithm, then the following holds true: 
$$d^\prime(C) \geq d^*(C)$$
In other words, the No-Skip base layer forward algorithm  achieves the minimum stall duration.
\label{lemma: noSkipLemma1}
\end{lemma}
\begin{proof}
Proof is provided in  Appendix~\ref{apdx:lemma1Noskip}.
\end{proof}
From Lemma \ref{lemma: noSkipLemma1}, we note that No-Skip forward algorithm would finish playing all chunks at their earliest time. Since all the chunks are obtained at the base layer quality and there is a minimum number of stalls, we note that the objective function is optimized for any $\beta\ge 1$ when only base layer is considered. When running base layer backward algorithm, the deadlines of the chunks are shifted to the last possibilities which gives the maximum flexibility of obtaining higher layers of chunks before their deadlines.

Having shown the result for the base layer and having determined the deadline for the last chunk, the rest of the algorithm is similar to the skip version where only the weighted quality need to be considered (the stall time is already found). Thus, the optimality result as described in the following Theorem holds, where the proof follows the same lines as described for the skip version theorem.

%

\begin{theorem}
If $z_{m,i}^*$ is the feasible size of every layer $m \leq M$ of chunk $i$ that is found by running No-Skip Layered Bin Packing Adaptive Algorithm, and $z_{m,i}^\prime$ is a feasible size that is found by any other feasible algorithm for the same stall duration, then the following holds for $0 <\gamma < 1$,  (\ref{basic_gamma_1}), $\beta \ge 1$, and  (\ref{lambda_cond}):
\begin{align*}
& \sum_{m=0}^M \gamma^m\sum_{i=1}^{C}\beta^i Z_{m,i}^\prime \leq \sum_{m=0}^{M} \gamma^m\sum_{i=1}^{C}\beta^i Z_{m,i}^*
\label{equ:eq1_lemma1}
\end{align*}
In other words, No-Skip Layered Bin Packing Adaptive Algorithm achieves the optimal solution of the optimization problem~(\ref{equ:eq2}-\ref{equ:c7eq2}). 
\label{thm:noskip}
\end{theorem}
\begin{proof}
Proof is provided in  Appendix~\ref{apdx:lthem2Proof}.
\end{proof}

The No-Skip scheme faces the same challenges described in~\S\ref{sec:bw_err}: short bandwidth prediction in the distant future and inaccurate bandwidth prediction, and they are handled the same way described in section~\S\ref{sec:bw_err}.

\section{EVALUATION}
\label{sec:eval}


In this section, we evaluate our algorithms (LBP) using both simulation and emulation.
Simulation allows us to explore a wide spectrum of the parameter space.
%
%
%
We then implemented a TCP/IP-based emulation testbed to compare its performance with simulation and to measure the runtime overhead in~\S\ref{sec:eval_emulation}.

%



\subsection{Simulation Parameters}

\begin{table}[htb]
	  \vspace{-.1in}
  \centering
  \caption{SVC encoding bitrates used in our evaluation}
  \begin{tabular}{|c|cccc|} \hline
    playback layer & BL & EL1 & EL2 & EL3 \\ \hline
   {nominal Cumulative} rate (Mbps) & 0.6 & 0.99 & 1.5 & 2.075 \\ \hline
  \end{tabular}
  \label{tab : svc_rates}
  \vspace{-.1in}
\end{table}

{\bf Simulation Setup.} To make our simulation realistic, we choose the SVC encoding rates of an SVC encoded video ``Big Buck Bunny", which is published in~\cite{SVCDataset}. It consists of 299 chunks (14315 frames), and the chunk duration is 2 seconds (48 frames and the frame rate of this video is 24fps).  The video is SVC encoded into one base layer and three enhancement layers. Table~\ref{tab : svc_rates} shows the cumulative nominal rates of each of the layers. The exact rate of every chunk might be different since the video is VBR encoded.  In the table, ``BL'' and ``EL$_i$'' refer to the base layer and the cumulative (up to) $i$th enhancement layer size, respectively. For example, the exact size of the $i$th enhancement layer is equal to EL$_i$-EL$_{(i-1)}$. 

For all schemes (both the baseline approaches and our algorithms),
we assume a playback buffer of 10 seconds ($B_{m}=10s$) for the skip version and 2 minutes for the No-Skip version, and a startup delay of 5 seconds. We will systematically study the impact of different algorithm parameters, including prediction accuracy, prediction window size, and playback buffer size  in Appendix \ref{sec:eval_para}. Finally, for all the variants of our algorithms with short prediction ($W \leq 20s$), we choose the lower buffer threshold to be half of the maximum buffer occupancy ($B_{min}=B_{m}/2$).  {When the buffer is less than $B_{min}$, we drop the highest layer that was decided to be fetched (unless the decision is fetching only the base layer). We still run the optimization problem, collect the layer size decisions, but we decrement the number of layers by 1 if enhancement layers are decided to be fetched. This helps being optimistic when the buffer is running low since the algorithm with short prediction have limited knowledge of the bandwidth ahead.
All reported results are based on the 50 diverse bandwidth traces described next.}

\begin{figure}
	\vspace{-.2in}
	\includegraphics[trim=0in 0in 0in 0in, clip,width=0.48\textwidth]{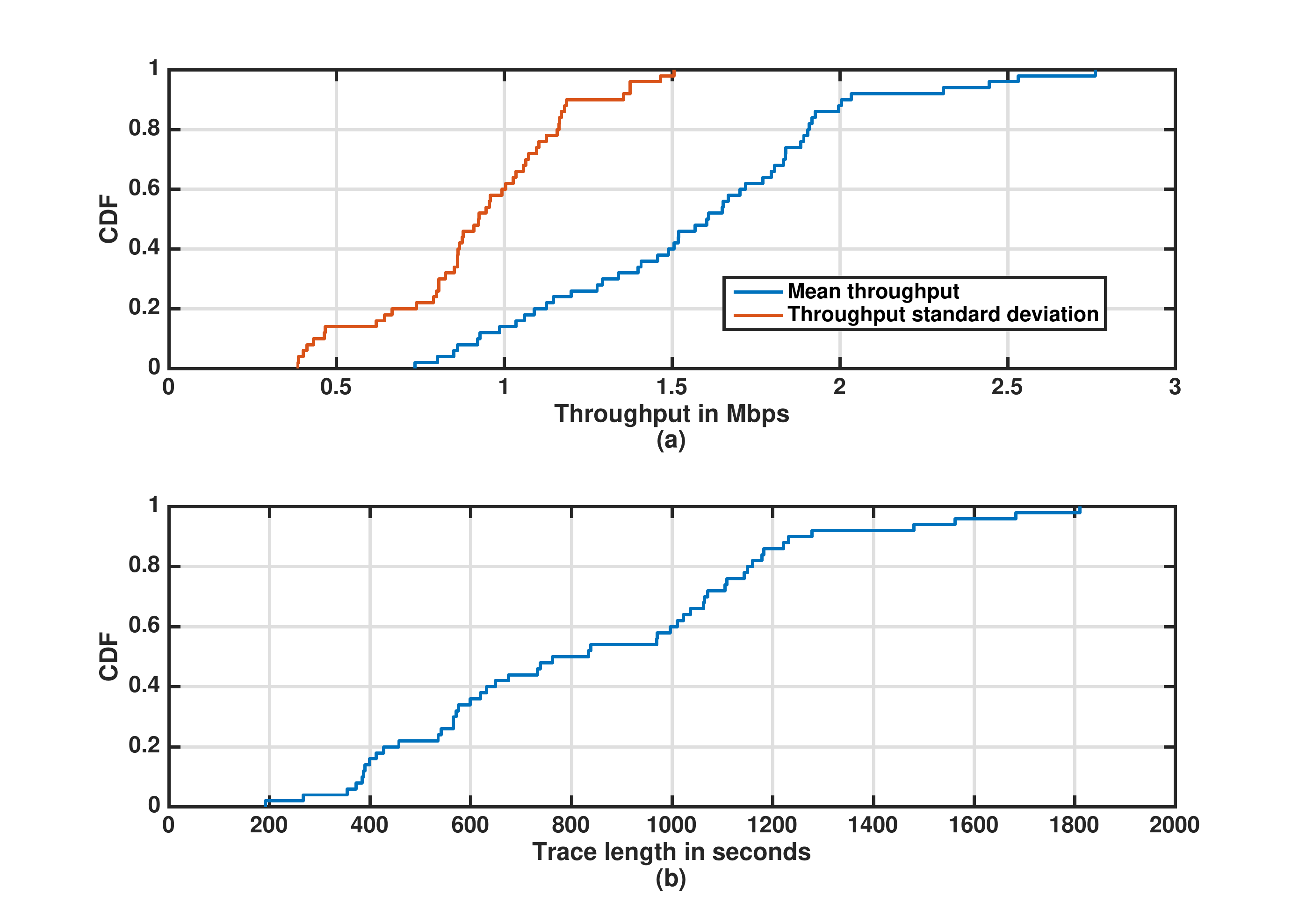}	
	\vspace{-.1in}
	\caption{Statistics of the bandwidth traces: (a) mean and standard deviation of each trace's throughput, and (b) trace length, across the 50 traces. \feng{Mark (a) and (b) in the plots}}
	\label{fig : bwStatV2}
	\vspace{-.1in}
\end{figure}

{\bf Bandwidth traces.} For bandwidth traces, we used the dataset in~\cite{riiser2013commute}, which consists of continuous 1-second measurement of video streaming throughput of a moving device in Telenor's
3G/HSDPA mobile network in Norway. The dataset contains 86 bandwidth profiles (traces) for different transportation types including bus, car, train, metro, tram, and ferry. We exclude traces with either very high or low bandwidth since in both cases the streaming strategies are trivial (fetching all layers and only base layers, respectively).
We then ended up having 50 traces whose key statistics are plotted in Fig.~\ref{fig : bwStatV2}.
Overall the traces are highly diverse, with lengths varying from 3 to 30 minutes. We note that since the ``Big Buck Bunny" is 598s. The video is re-started for long traces and cut at the end of the trace for short traces.

The average throughput across the traces varies from 0.7Mbps to 2.7 Mbps, with the median being 1.6 Mbps. In
each trace, the instantaneous throughput is also highly variable, with the average standard deviation across traces being 0.9 Mbps. 

{\bf Bandwidth Prediction.}
We consider two different techniques for bandwidth prediction. First is a harmonic mean based prediction in which the harmonic mean of the bandwidth of the last 5 seconds is used as a predictor of the bandwidth for the next 20 seconds. We refer to our algorithm with harmonic mean based prediction by HM. Second, we assume crowd sourced prediction, and  a combination of prediction window size with prediction error percentages. Longer prediction window comes with the cost of higher prediction error. For example we use $(10,25\%)$ to refer to the prediction window ($W$) of 10 seconds and the prediction error $pe$ of 25\%. In our simulation, the predicted bandwidth is computed by multiplying the actual value in the bandwidth trace (the ground truth)  by $1+e$ where $e$ is uniformly drawn from $[-pe, pe]$ (based on our findings in Appendix \ref{sec:predict}, the prediction error tends to have a mean of 0 in the long run).   
For skip version (real time streaming), we evaluated our algorithm in case of $(10,25\%)$ and $(20,50\%)$ since chunks beyond 20 seconds ahead might not be available yet. However, for the No-Skip version (non-real time streaming), we considered $(20,50\%)$ and $(100,60\%)$. We also include the offline scheme \ie $(\infty,0)$, for comparison. It corresponds to the performance upper bound for an online algorithm, which is given by our offline algorithm.

\subsection{Skip Based Streaming}
\label{sec:eval_skip}
We compare our skip-based streaming algorithm (\S\ref{skipalgo}) with three baseline algorithms with different aggressiveness levels. Baseline 1 is a conservative algorithm performing ``horizontal scan'' by first trying to fetch the base layer of all chunks up to the full buffer.
If there is spare bandwidth and the playout buffer is not full, the algorithm will fetch the first enhancement layer of buffered chunks that can be received before their playback deadline. If the bandwidth still permits, the algorithm will fetch the second enhancement layer in the same manner.
%
Baseline 2 instead aggressively performs ``vertical scan'', it fetches all layers of the next chunk before fetching the future chunks. 
Baseline 3 is a hybrid approach combining Baseline 1 and 2. It first (vertically) fetches all layers of the next chunk and if there is still available bandwidth, it subsequently (horizontally) fetches the base layer of all later chunks before proceeding to their higher layers.

%

%

\begin{figure}
\includegraphics[trim=0in .1in 0in 0in, clip,  width=0.48\textwidth]{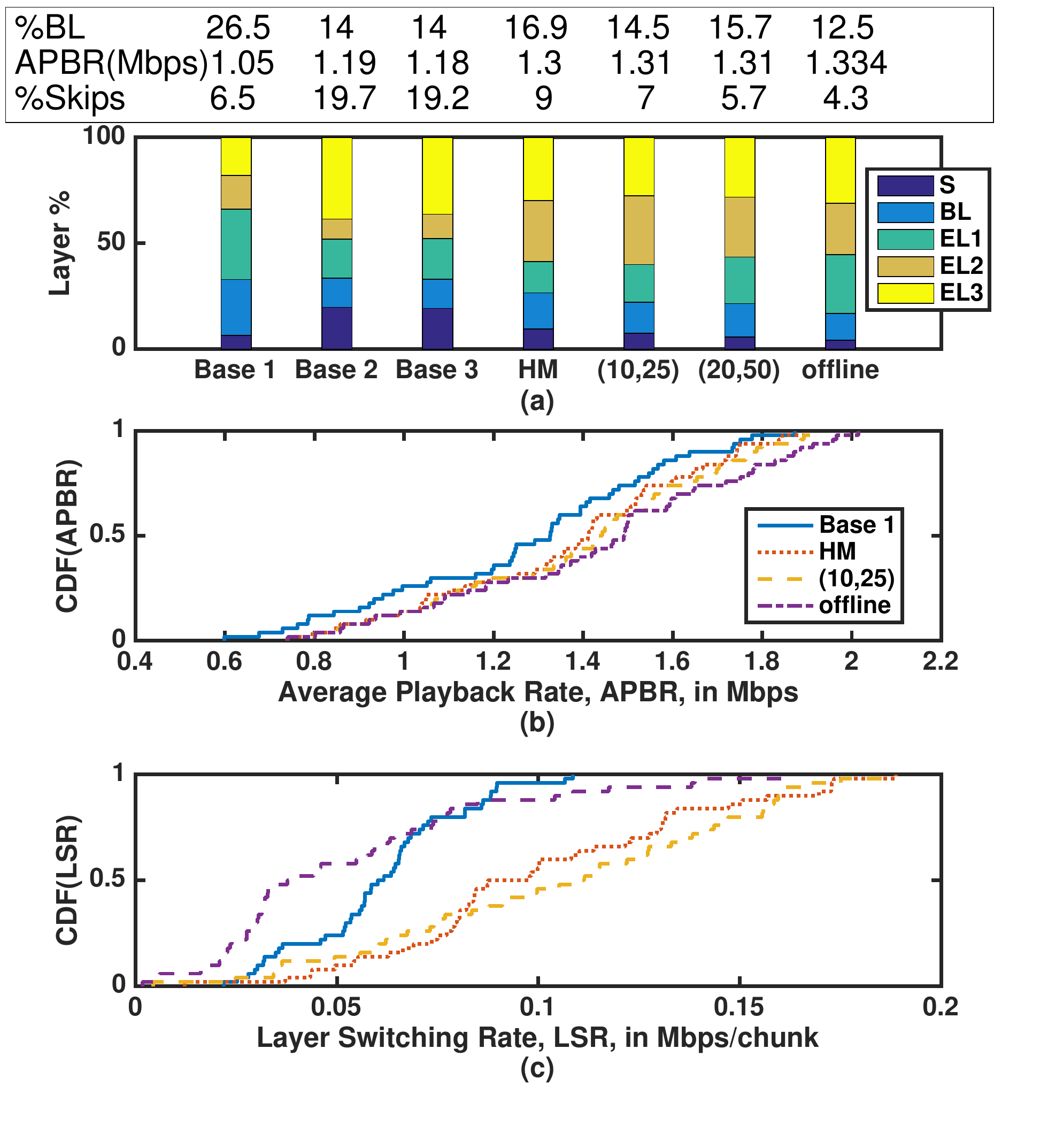}
 \caption{Skip based streaming results for different schemes: (a) layer distribution, (b) average playback rate, and (c) layer switching rate.}
 \label{fig : BComp}
\vspace{-.1in}
\end{figure}

We compare the above three baseline approaches with three representative configurations of our proposed online LBP algorithm.
They are referred to as HM (harmonic mean based prediction), $(10,25\%)$, and $(20,50\%)$. Moreover, we include our offline algorithm which has a perfect bandwidth prediction for the whole period of the video.

 The results are shown in the three subplots of Fig.~\ref{fig : BComp}. Fig.~\ref{fig : BComp}-a plots the
breakdown of the highest fetched layers of each chunk (``S'' refers to skipped chunks).
For example, for Baseline 1, $26.5$\% of chunks are fetched only at the base layer quality (shown in light blue).
The average playback rate (across all 50 traces) for each scheme is also marked in the plot.
As shown, our schemes significantly outperform the three baseline algorithms by fetching more chunks at higher layers with fewer skips. Even when the prediction window is as short as 10 seconds, our scheme incurs negligible skips compared to Baseline 2 and 3, and yields an average playback bitrate that is $\sim$25\% higher than Baseline 1.
As the prediction window increases (\ie $W=20s$ and $pe=50\%$), the layer distribution becomes very close to the offline scheme.

Fig.~\ref{fig : BComp}-b plots the CDF of the average playback rate of all the schemes across all traces.
As shown, even with a prediction window of as short as 10 seconds, 
our online scheme achieves playback rates that is the closest to those achieved by the offline scheme across the 50 traces. One more interesting observation from Fig.~\ref{fig : BComp}-b is that both variants of our algorithm (HM, and ($10$,$25\%$)) outperform Baseline 1 in terms of average playback rate in every  bandwidth trace.
Also note that although Baseline 2 and 3 achieve higher playback rates than Baseline 1, they suffer from a large number of skips as shown in Fig.~\ref{fig : BComp}-a.

Fig.~\ref{fig : BComp}-c plots for each algorithm the distribution of the layer switching rates (LSR),
which is defined as $\frac{1}{C*L}\sum_{i=2}^C |X(i)-X(i-1)|$ where $C$ is the number of chunks, $L$ is the chunk duration, and $X(i)$ is the size of chunk $i$ (up to its fetched layer). Intuitively, LSR quantifies the frequency of the playback rate change, and ideally should be minimized.
%
%
Baseline 1, behaves very conservatively by first fetching the base layer for all chunks up to full buffer. Therefore it has lower layer switching rates  at the cost of lower playback rates. Our algorithms instead
achieve reasonably low layer switching rates while being able to stream at the highest possible rate with no skips.

{ We note that larger prediction windows can lead to better decisions even if the prediction has higher error. As long as the bandwidth prediction is unbiased, we see that higher prediction errors can be tolerated. Appendix  \ref{sec:predict} shows that crowdsourcing-based prediction is an unbiased predictor of the future bandwidth. Moreover, more results about the effect of the prediction error on the proposed algorithm are described in  Appendix \ref{sec:eval_para}. Further, we show that the computational overhead of the proposed approach is low, as described in Appendix \ref{apdx:comp_time}.
}



\subsection{No-Skip Based Streaming}
\label{sec:eval_noskip}

\begin{figure}
\includegraphics[trim=0in 0in 0in 0.1in, clip, width=0.48\textwidth]{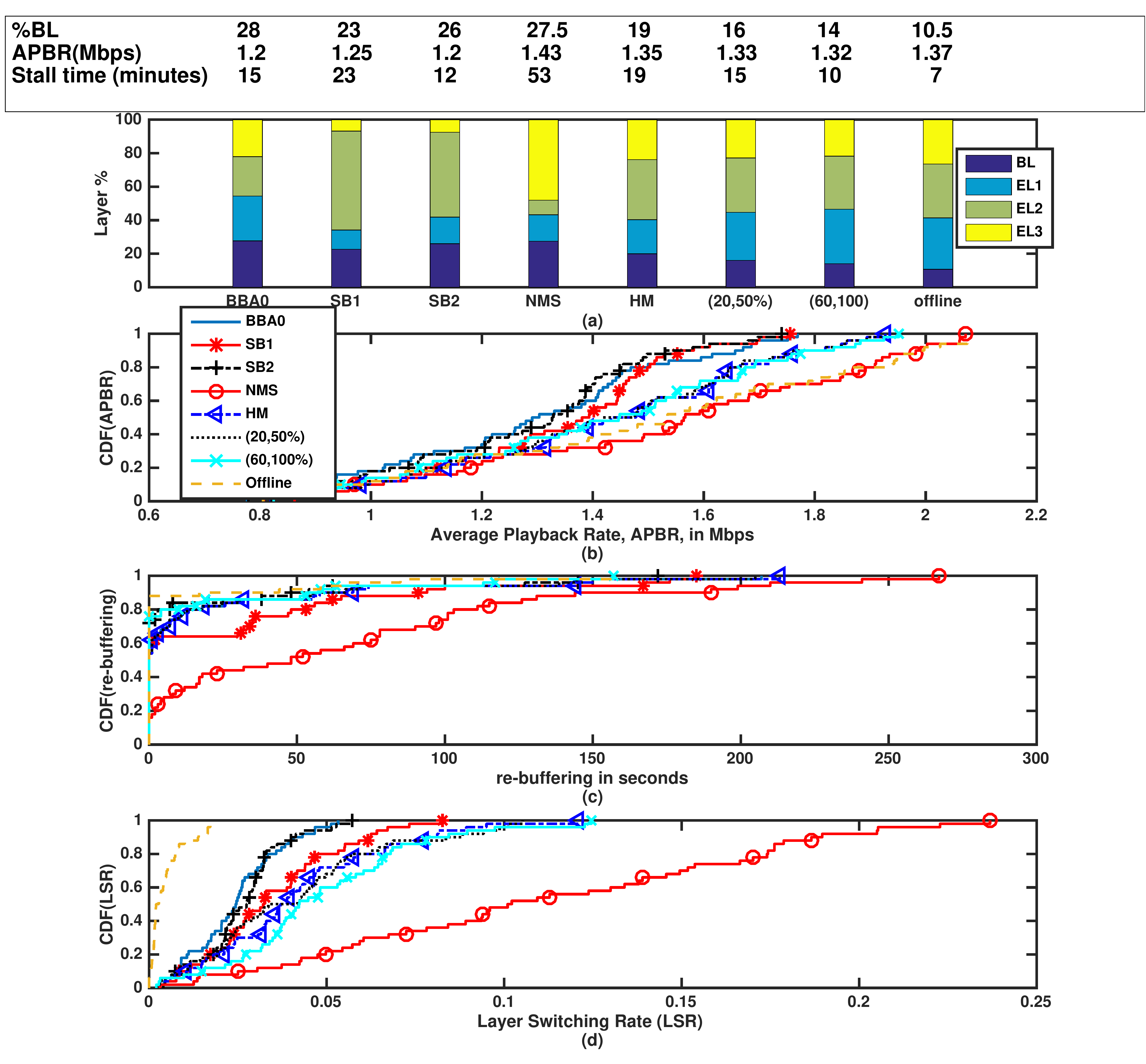}	
 \vspace{-.1in}
 \caption{
 No-Skip based streaming results for different schemes: (a) layer distribution, (b) average playback rate, (c) total rebuffering time, and (d) layer switching rate. \feng{Change ``Stall time in minutes'' to ``Stall time (min)''.}}
  \label{fig:Comp_ns}
  \vspace{-.1in}
 \end{figure}

We now evaluate the no skip based algorithm. We compare it with three 
state-of-the-art algorithms: buffer-based algorithm (BBA) proposed by Netflix~\cite{BBA}, 
Naive port of Microsoft's Smooth Streaming algorithm for SVC~\cite{SVCMetrics}, and a state-of-the-art slope-based SVC streaming approach~\cite{Andelin12}.
To ensure apple-to-apple comparisons, we adopt the same parameter configuration (2-minute buffer size and 1-second chunk size) and apply the algorithms to all our 50 traces. Before describing the results, we first provide an overview of the three algorithms we compare our approach with.


{\bf Netflix Buffer-based Approach (BBA~\cite{BBA})} adjusts the streaming quality based on the playout buffer occupancy.
Specifically,
it is configured with lower and upper buffer thresholds. If the buffer occupancy is lower (higher) than the lower (higher) threshold, chunks are fetched at the lowest (highest) quality;
if the buffer occupancy lies in between, the buffer-rate relationship is determined by a pre-defined step function.
We use 40 and 80 seconds as the lower and upper thresholds. The quality levels are specified 
in terms of the SVC layers (\eg ``the highest quality'' means up to the highest layer).


{\bf Naive port of Microsoft Smooth Streaming for SVC~\cite{SVCMetrics} (NMS)}
employs a combination of buffer and instantaneous bandwidth estimation for rate adaptation.
NMS is similar to BBA in that it also leverages the buffer occupancy level to determine the strategy.
The difference, however, is that 
it also employs the instantaneous bandwidth estimation (as opposed to the long-term
network quality prediction we use) to guide rate adaptation. As a result, for example, it can fetch high-layer chunks
without waiting for the buffer level reaching the threshold as is the case for BBA.

{\bf Slope-based SVC Streaming~\cite{Andelin12}} \feng{reference missing} takes the advantage of SVC over AVC. It can download the base layer of a new chunk or increase the quality of a previously downloaded (but not yet played) chunk by downloading its enhancement layers.
This is achieved by defining a slope function: the steeper the slope, the more backfilling will be chosen over prefetching. 
Following the original paper's recommendations, we empirically choose 2 slope levels (SB1: -7\%, and SB2: -40\%).
We verified that these two settings provide good results compared to other slope configurations (\eg 
going steeper than SB1 causes longer stall duration and going flatter than SB2 makes the playback rate lower).


\textbf{The results} are shown in four subplots in Fig.~\ref{fig:Comp_ns}. Fig.~\ref{fig:Comp_ns}-a plots the
layer breakdown.
The average playback rate and the total rebuffering time (across all 50 traces) for each scheme are also marked.
As shown, in terms of rebuffering time, our online schemes with crowd sourced bandwidth prediction achieve the lowest stall duration even when the prediction window is as short as 20 seconds ahead. On other hand, NMS performs poorly in terms of avoiding stalls since It runs into almost an hour of stalls (53 minutes). Moreover, all variants of our online algorithm including HM significantly outperform other algorithms in fetching higher layers. For example, (20,50\%) fetches only $16\%$ of the chunks at BL quality which is 57\%, 70\%, 62\%, and 58\% fewer then BBA0, SB1, SB2, and NMS respectively. Also, as the prediction window increases, the layer distribution becomes closer to the offline scheme, with the shortest stall duration incurred.
%
%
Fig.~\ref{fig:Comp_ns}-b and Fig.~\ref{fig:Comp_ns}-c plot for each algorithm the distribution of the (per trace) average playback rate and the stall duration across all traces. The results are consistent with our findings from Fig.~\ref{fig:Comp_ns}-a: our scheme achieves high playback rate that is the closest to the very optimistic algorithms (\eg NMS) while incurring stalls that are as infrequent as the very conservative algorithms (\eg SB3 and BBA). Thus, it is clearly shown that our algorithm is maintaining a good trade-off between minimizing the stall duration and maximizing the average playback rate. 
Fig.~\ref{fig:Comp_ns}-d plots for each algorithm the distribution of the layer switching rates (LSR, defined in~\S\ref{sec:eval_skip}). Similar to the skip based scenario, our schemes achieve much lower LSR compared to
the aggressive approach (\eg NMS). The LSR can further be reduced but at the cost of reduced playback rate.

{ To conclude this section, we would like to point out the key points behind achieving better performance for our algorithm as compared to the baselines. First, incorporating chunk deadlines, bandwidth prediction, and buffer constraint into the optimization problem yields a better decision per chunk. Moreover, favoring the later chunks helps the algorithm avoid being overly optimistic now at the cost of running into skips later on. Finally, re-considering the decisions after the download of every chunk with the new updated bandwidth prediction helps make the algorithm self-adaptive and more dynamically adjustable to the network changes. The low complexity of the algorithm allows for re-running the algorithm and changing decisions on the fly.}




\subsection{Emulation over TCP/IP Network}
\label{sec:eval_emulation}

To complement our simulation results, we have built an emulation testbed using C++ (about 1000 LoC) on Linux.
The testbed consists of a client and a server. All streaming logics described in~\S\ref{sec:alg} are implemented on the client side, which fetches synthetic chunks from the server over a persistent TCP connection. We deploy our emulation testbed between a commodity laptop and a server inter-connected using high-speed Ethernet (1Gbps link and 1ms RTT).
We use \texttt{Dummynet}~\cite{dummynet} on the client side to replay a bandwidth profile by dynamically changing the available bandwidth every one second.
We also use the Linux \texttt{tc} tool to inject additional latency between the client and server.

\begin{table}[htb]
	\centering
	\caption{LTE bandwidth traces}
	\begin{tabular}{|c|cccccc|} \hline
		Trace No. & 1 & 2 & 3 & 4 & 5 & 6 \\ \hline
		Average rate (Mbps) & 5.05 & 6.95 & 5.9 & 6.14&5.3&6.8 \\ \hline
		Standard deviation (Mbps) & 4.3 & 6.65 & 4.7 & 5.25&3.84&7.02 \\ \hline
	\end{tabular}
	\label{tab : lte_traces}
	\vspace{-.1in}
\end{table}

We next run the emulation experiment using six bandwidth traces, each of length 15-minutes. These traces were collected on an LTE network on different drive routes (as described in Appendix \ref{sec:predict}). Table \ref{tab : lte_traces} shows the statistics of the bandwidth traces, and since the bandwidth of the traces are high, we used the following cumulative SVC rates, $1.5Mbps$ (BL), $2.75Mbps$ ($EL_1$), $4.8Mbps$ ($EL_2$), $7.8Mbps$ ($EL_3$) ~\cite{SVCDataset}. We configure the end-to-end RTT to be 60ms, which roughly corresponds to the last-mile latency in today's LTE networks. Meanwhile, we run the same bandwidth traces under identical settings using the simulation approach. Since all traces confirm similar behavior, we explain the results of one bandwidth trace, so we can have both the quality CDF and the playback quality over time. 

\begin{figure}
\centering
	\includegraphics[trim=1.2in 0in 1.4in .1in, clip, width=0.48\textwidth]{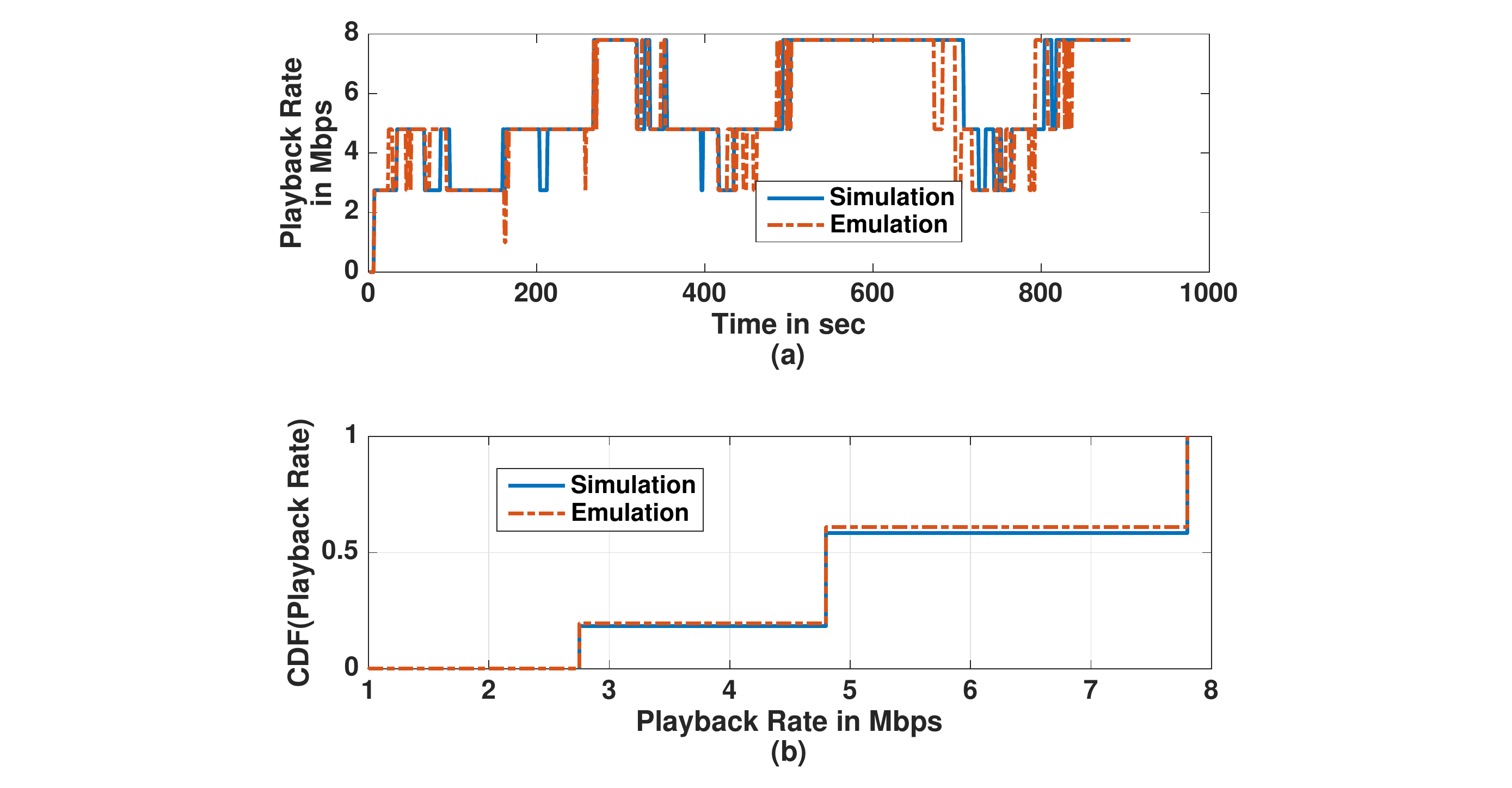}
	\vspace{-.1in}	
	\caption{Emulation vs simulation: (a) playback bitrate over time, (b) chunk quality distribution. \feng{Reduce the space between the two plots. Remove ``Empirical CDF''.}}
	\label{fig:emu}
	\vspace{-.1in}
\end{figure}

Fig.~\ref{fig:emu}-a compares the simulation and emulation results in terms of the qualities of fetched chunks, and Fig.~\ref{fig:emu}-b compares the chunk quality distribution. As shown, the simulation and emulation results well cross-validate each other. Their slight difference in Fig.~\ref{fig:emu}-a is mainly caused by the TCP behavior (\eg slow start after idle) that may underutilize the available bandwidth. 

%
%
%
%

\section{Conclusions and Future work}

We formulated the SVC rate adaptation problem as a non-convex optimization problem that has an objective of minimizing the skip/stall duration as the first priority, maximize the average playback as the second priority, and minimize the quality switching rate as the last priority. We develop LBP (Layered Bin Packing Adaptive Algorithm), a low complexity algorithm that is shown to solve the problem optimally in polynomial time. Therefore, offline LBP algorithm that uses perfect prediction of the bandwidth for the whole period of the video  provides a theoretic upper bound. Moreover, an online LBP that is based on sliding window and solves the optimization problem for few chunks ahead was proposed for the more practical scenarios in which the bandwidth is predicted for short time ahead and has prediction errors. The results indicate that LBP 
is robust to prediction errors, and works well with short prediction windows. It outperforms existing streaming approaches 
by improving key QoE metrics. Finally,  LBP incurs low runtime overhead due to its linear complexity. 

Extending the results to consider streaming over multiple paths with link preferences is an interesting problem, and is being considered by the authors in \cite{anisspcom,multipath,groupcast}.

%

\bibliographystyle{IEEEtran}
\bibliography{refs,bib}
\begin{IEEEbiography}[{\includegraphics[width=1in,height=1.25in,clip,keepaspectratio]{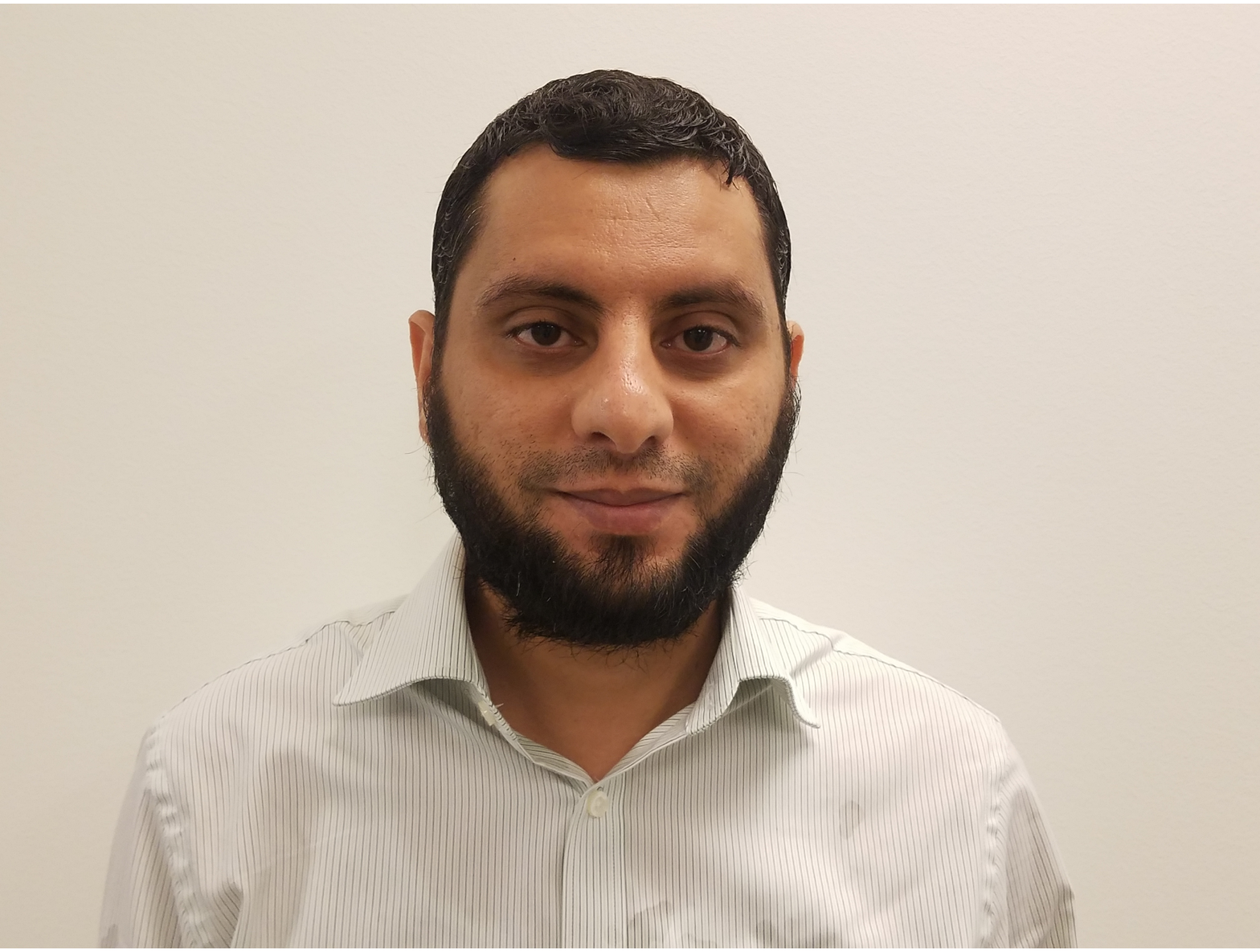}}]{Anis Elgabli} received the B.S. degree in electrical and electronic engineering from University of Tripoli, Tripoli, Libya in 2004. Further, he received the M.Eng. degree from UKM, Kajang, Malaysia in 2006, M.S. and Ph.D. degrees in 2015 and 2018, respectively from Purdue University,  IN, USA, all in Electrical and Computer Engineering.
	His research interest is in applying optimization techniques in networking and communication systems. He was the recipient of the 2018 Infocom Workshop HotPOST Best Paper Award.  
\end{IEEEbiography}

\begin{IEEEbiography}[{\includegraphics[trim=.5in 0in .5in 0in, clip,width=1in,height=1.25in,clip,keepaspectratio]{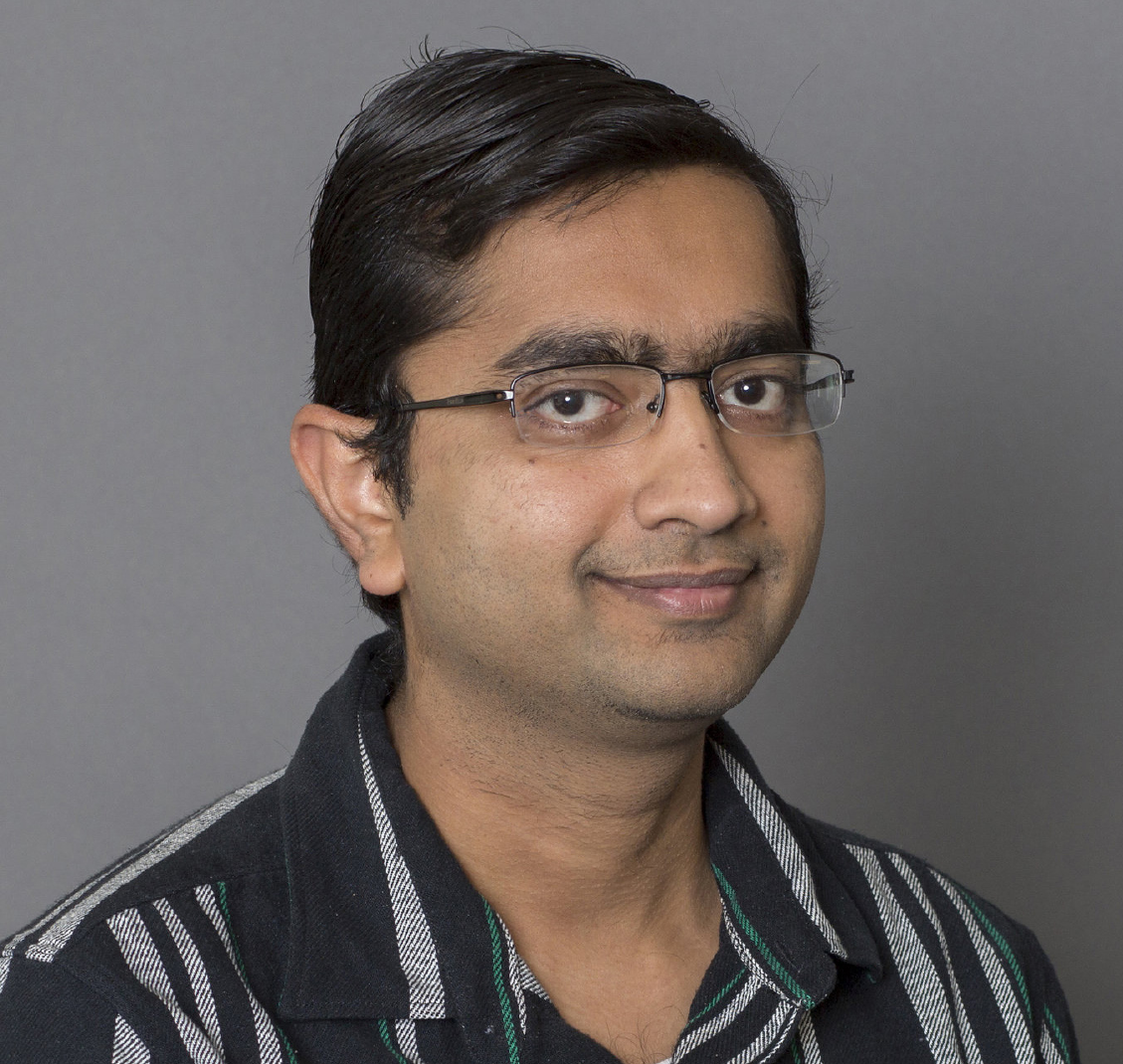}}]{Vaneet Aggarwal (S'08 - M'11 - SM'15)}
received the B.Tech. degree in 2005 from the Indian Institute of Technology, Kanpur, India, and the M.A. and Ph.D. degrees in 2007 and 2010, respectively from Princeton University, Princeton, NJ, USA, all in Electrical Engineering.

He is currently an Assistant Professor at Purdue University, West Lafayette, IN. He is also a VAJRA Adjunct Professor at IISc Bangalore. Prior to this, he was a Senior Member of Technical Staff Research at AT\&T Labs-Research, NJ (2010-2014), and an Adjunct Assistant Professor at Columbia University, NY (2013-2014). His current research interests are in communications and networking, video streaming, cloud computing, and machine learning. 

Dr. Aggarwal is on the editorial board of the {\it IEEE Transactions on Communications} and the {\it IEEE Transactions on Green Communications and Networking.} He was the recipient of Princeton University's Porter Ogden Jacobus Honorific Fellowship in 2009, the AT\&T Key Contributor award in 2013, the AT\&T Vice President Excellence Award in 2012, and the AT\&T Senior Vice President Excellence Award in 2014. He was also the recipient of the 2017 Jack Neubauer Memorial Award, recognizing the Best Systems Paper published in the IEEE Transactions on Vehicular Technology and the 2018 Infocom Workshop HotPOST Best Paper Award.  
\end{IEEEbiography}

\begin{IEEEbiography}[{\includegraphics[width=1in,height=1.25in,keepaspectratio]{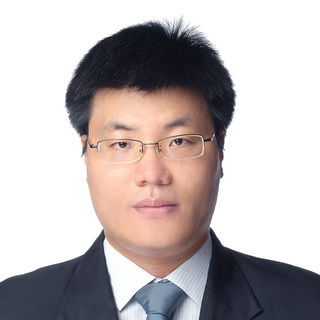}}]{Shuai Hao}  received the B.Comp. (First-class honors) and M.S. degrees, in 2005 and 2006, respectively, from the National University of Singapore, Singapore, and the Ph.D. degree in 2014 from the University of Southern California, Los Angeles, CA, USA, all in Computer Science. 
	
	He is currently a Senior Inventive Scientist at AT\&T Labs - Research, NJ, which he joined in 2014. Prior to this, he was a Research Assistant (2012-2014) and an Annenberg Fellow (2008-2012) at University of Southern California, and a Research Staff at National University of Singapore (2006-2008). His research interests include networked systems, mobile computing, sensor network and IoT, and video streaming.
	
	Dr. Hao has published in ACM MobiSys, CoNEXT, ICSE, and IEEE INFOCOM. His work on reducing web latency was published in ACM SIGCOMM’13 and received the Applied Networking Research Prize, awarded by the Internet Research Task Force at the IETF 91 meeting in 2014.
\end{IEEEbiography}

\begin{IEEEbiography}[{\includegraphics[width=1in,height=1.25in,keepaspectratio]{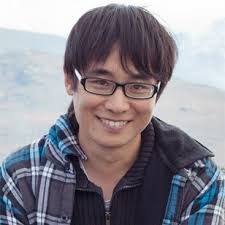}}]{Feng Qian} is an assistant professor in the Computer Science Department at Indiana University Bloomington. His research interests cover the broad areas of mobile systems, VR/AR, computer networking, and system security. He obtained his Ph.D. at the University of Michigan, and his bachelor degree at Shanghai Jiao Tong University.
\end{IEEEbiography}

\begin{IEEEbiography}[{\includegraphics[width=1in,height=1.25in,clip,keepaspectratio]{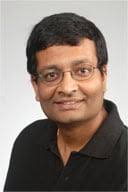}}]{Subhabrata Sen (M'01-SM'14-F'16)} received the Ph.D. degree in computer science from the University of Massachusetts, Amherst, MA, USA, in 2001. He is currently a Lead Scientist at AT\&T Labs-Research, where he has been since 2001. His research interests include Internet technologies and applications, IP network management, application and network performance, video streaming, cross- layer interactions and optimizations for cellular net- works, network measurements, and traffic analysis. He has co-authored 96 peer-reviewed 
	research articles in leading journals, conferences and workshops, and holds 62 awarded patents. 
	
	Dr. Sen was a recipient of the AT\&T Science and Technology Medal, the AT\&T Labs President Excellence Award, and the AT\&T CTO Innovation Award. He is a co-inventor of the widely used open-source Application Resource Optimizer tool for analyzing cellular friendliness of mobile app designs that earned top industry honors, including the American Business Awards Tech Innovation of the Year Gold Stevie Award, in 2013. He served in the past as an Editor of the IEEE/ACM Transactions on Networking.
 \end{IEEEbiography}
\clearpage

\setcounter{page}{1}

\appendices
\section{Why using SVC Encoding?}
\label{sec:whySVC}

As of today, Adaptive-SVC streaming has not yet registered wide deployment in commercial video platforms likely due to its complexity and a lack of use cases. Also as expected, much less academic research has been conducted on Adaptive-SVC streaming as compared to ABR streaming which is the streaming logic of AVC encoded videos. Nevertheless, we believe that SVC's unique encoding scheme may help provide better adaptiveness and scalability than the state-of-the-art, in particular under challenging network conditions such as mobility.
The key reason is that \emph{SVC allows the quality of a video chunk to change incrementally.} Consider the following common scenarios.

\BULLET In ABR streaming, different versions of a chunk are independent. If a chunk is not fully downloaded before its playback deadline, a stall will occur. 
This issue can be easily mitigated by SVC:
if layer $i+1$ is not fully downloaded at the chunk playback deadline, the chunk is still playable at a lower quality of up to layer $i$. SVC encoding has indeed been shown to be able to cope much better adjustability with compare to AVC encoding with highly variable bandwidth~\cite{SVCMetrics}, which is one of the key characteristics of cellular networks.

\BULLET Assume a player switches to a higher bitrate during a playback. This can be triggered either manually or automatically by the rate adaptation algorithm. Then consider those chunks that are already downloaded to the player's buffer. For ABR streaming, those buffered low-quality chunks are either played or discarded if the player replaces them with high-quality chunks. As a result, the actual quality switch either happens later or happens immediately with bandwidth waste incurred. This dilemma can be elegantly addressed by SVC: the player can reuse the buffered chunks by augmenting them using enhancement layers.

\BULLET SVC can also make video caching more efficient. In ABR streaming, a caching proxy needs to cache all the \emph{versions} of an AVC encoded chunk (\eg up to 8 versions for a YouTube video). For an SVC chunk, the proxy only needs to cache all its \emph{layers}, whose total size is equivalent to that of a single AVC chunk at the highest quality plus the SVC encoding overhead which is typically around 10\%.

\BULLET Adaptive-SVC streaming enables many other novel use cases. For example, many content providers (\eg Facebook) today embed videos into web pages. An emerging type of page-embedded video is \emph{auto-play video}, which is automatically streamed and played when the video player window appears on the screen as a user scrolls down a ``long'' web page.
Usually auto-play videos are automatically streamed at a lower quality. When the user clicks/touches the video player, the player will turn on the sound and continue playing the video at a higher quality. The above scheme suffers from similar bandwidth inefficiency as described in the second point, which can be similarly alleviated by SVC.  Further, it can help save storage space for caching content closer to the client. 

\begin{thisnote}
	We note that the authors of \cite{muller2012using} provided SVC extensions of the AVC standard into MPEG-DASH standard using byte-range requests.
\end{thisnote}

\section{Predictability of Available Network Bandwidth}
\label{sec:predict}
Video streaming is highly bandwidth-intensive.
Our adaptive SVC streaming algorithms leverage the knowledge of future available network bandwidth.
Now we investigate how predictable the future bandwidth is in state-of-the-art commercial LTE networks.
We collected network bandwidth traces over a period of three months for one commute route between an employee's home and work.
The route is 14.2 miles long and comprised of half local road and half highway road, with respective speed limits of 40mph and 65mph.
We used a Samsung Galaxy S3 phone for bandwidth collection in a major U.S. cellular carrier's LTE network.
We also developed an Android app to continuously measure the available TCP throughput between the phone and a remote server.
Note that we made sure the remote server has enough available bandwidth to the Internet so that the measured throughput at phone can best reflect the overall end-to-end available bandwidth.
We turned on GPS and collected the location trace every 10 meters.
After each experiment, we will get two timestamped traces: available bandwidth and GPS location.
Although we have tried to start the experiment at the same time and to follow the same route, there are still factors, such as weather and traffic, that are out of our control and will make the trip take varying amounts of time.
To make fair analysis of traces collected over different days, we chose to use \emph{distance}-tagged bandwidth traces instead of the default timestamp-tagged ones.
To achieve that, we aligned the bandwidth and GPS traces with timestamp, calculated the distance between two consecutive GPS points, and associated the average bandwidth of the two points to that distance.
The traces are collected for a period of  over three months.

\begin{figure}[hbtp]
\begin{center}
 \includegraphics[trim=0in 0in 0in 0in, clip,width=0.5\textwidth, height=2.5in]{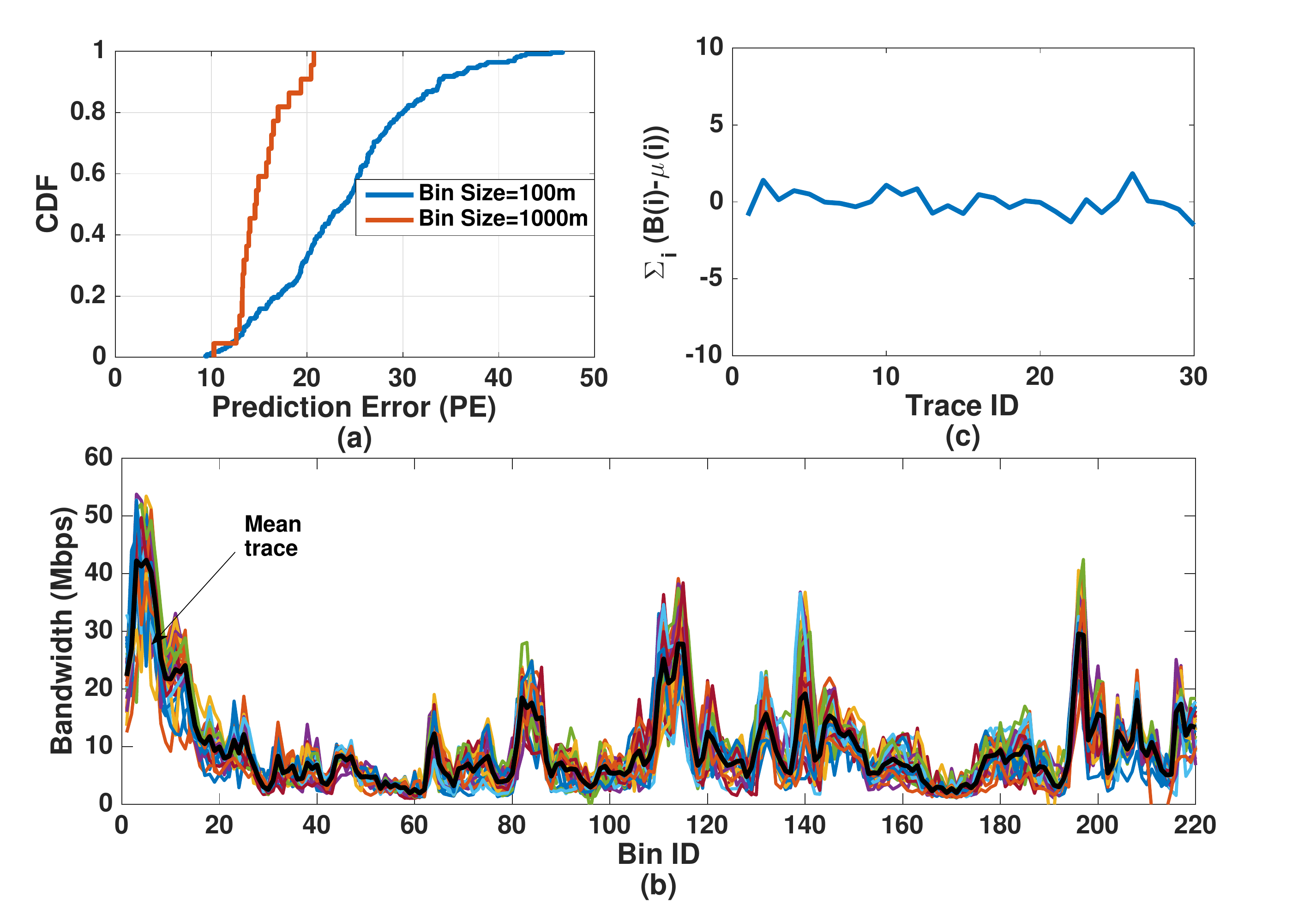}
 \vspace{-.1in}
\caption{A case study of bandwidth predicability in LTE networks: (a) distribution of prediction error across traces, (b) time series of individual traces and their mean trace, (c) signed differences (over all bins) between the mean trace and individual traces.}
\label{pred_analysis}
\end{center}
 \vspace{-.1in}
\end{figure}

Next, we present our results for the predictability analysis.
First, we divide each distance-tagged trace into bins with equal distance \emph{b} (\eg every $b$=100 meters).
Then we calculate the mean for bandwidth measurements that fall into the same bin and use that value to represent the average available bandwidth of the corresponding bin.
Finally, we perform statistical analysis for each bin across all traces.
%
%
Fig~\ref{pred_analysis}-a shows the cumulative distribution of the prediction error $pe$ across all traces.
We devise a simple prediction method as follows:
we use the \emph{mean trace} from the previous 30 days (\ie a trace with each bin's bandwidth being
calculated as the mean of the corresponding bins over the past 30 days) 
as the prediction of the bandwidth trace for the current day.
The $pe$ of a particular trace is then computed as $\sum_{i}|B(i)-\overline{B}(i)|/(B(i) \cdot N)$ where $B(i)$ and
$\overline{B}(i)$ are the $i$-th bin's bandwidth of the current trace (the groundtruth) and the mean trace (the predicted value), respectively, and $N$ is the number of bins.
%
%
As shown, when the bin size is 1000 meters (covering $\sim$37 seconds in time when the speed is 60mph),
$pe$ is less than 20\% for more than 90\% of the traces (median $pe$: 15\%).
As the bins become smaller, $pe$ increases accordingly as it is difficult to predict the instantaneous
bandwidth: when the bin size is 100 meters (covering 3.7 seconds at 60mph),
$pe$ is less than 35\% for more than 90\% of the traces (median $pe$: 25\%).
Overall the results indicate that historical information can well be leveraged to
predict future network conditions.
%

Another key observation we made is that 
the signed difference between the predicted bandwidth and the real bandwidth 
tends to have a mean of 0. In other words, an underestimation is very likely to be ``corrected'' by
an overestimation in the near future, and vice versa. This is visually illustrated in Fig~\ref{pred_analysis}-b,
which plots the time series of a set of 30 traces 
as well as that of their mean trace (the thick curve). 
More quantitatively, Fig~\ref{pred_analysis}-c plots $\sum_{i} \left( B(i)-\overline{B}(i) \right)$ for 30 traces.
As shown, the sum of signed difference over bins are indeed small. As Fig~~\ref{pred_analysis}-c shows, the max point is about 2Mb, so a difference of 2 Mb (about size of one chunk of a video) over a trace of about 20min length is negligible. Therefore, the mean of the traces for people travelled the same road recently is indeed an unbiased estimate of the current trace.

%

\begin{thisnote}
\section{Table of Notation}
\label{sec:notationTable}
In this section, we include a table of notation to make the tracking of variables easier.

 \begin{table}[htb]
	\vspace{-.1in}
	\centering
	\caption{Notation}
	\begin{thisnote}
		\resizebox{\columnwidth}{!}{%
			\begin{tabular}{|c|c|} \hline
				$s$&  Startup delay\\ \hline
				$B(j)$& Available bandwidth at time $j$\\ \hline
				$B_m$& Maximum buffer size\\ \hline
				$C$&Index of the last chunk in the video \\ \hline
				$d(i)$& Total stall duration before chunk $i$\\ \hline
				$deadline(i)$& Deadline of chunk $i$\\ \hline
				$L$&Chunk duration\\ \hline
				$r_n$& Rate of the $n$-th layer\\ \hline
				$X(i)$&  Size fetched of  chunk $i$\\ \hline
				$X_n$&  Cumulative size of the $n$-th layer of chunk $i$\\ \hline
				$Y_n$&  Size of the $n$-th layer of a video chunk\\ \hline
				$Z_{n,i}$&  Size decision of the $n$-th layer of chunk $i$, $Z_{n,i} \in \{0,Y_n\}$ \\ \hline
				$z_{n}(i,j)$& Size fetched of the $n$-th layer of chunk $i$ at time $j$ \\ \hline
				$t(i)$& Lower deadline of chunk $i$\\ \hline
				$a(i)$& Size fetched for chunk $i$ at its lower deadline time slot $t(i)$\\ \hline
				$e(j)$& Available bandwidth at time slot $j$ after all chunks are fetched\\ \hline
			\end{tabular}
		}
	\end{thisnote}
	\label{tab : notation}
	\vspace{-.1in}
\end{table}

\section{Illustrative Example to Explain the Intuition behind the Selected Objective Function, Eq.~\eqref{equ:eq1} }
\label{sec:formulationExample}

The objective function $\sum_{n=0}^{N}\gamma^n\sum_{i=1}^{C}\beta^i Z_{n,i}$ is a weighted sum of the layer sizes for different chunks. The weights are determined by two parameters $\gamma$ and $\beta$. The parameter $\gamma$ indicates that fetching the $n$-th layer of a chunk achieves a utility that is $0<\gamma<1$ times the utility that is achieved by fetching the $(n-1)$-th layer. The choice of $\gamma$ in \eqref{basic_gamma_1} implies that all the higher layers than layer $a$ have lower utility than a chunk at layer $a$ for all $a$. User's QoE is concave in the playback rate \cite{miller2017qoe}, so the higher layers contribute lower to the QoE as compared to the lower layers. For example, running into base layer skips degrades the QoE much more than skipping the first enhancement layer (E1) since when E1 is skipped the video chunk is still can be played. Thus, we assume that fetching higher layers have diminishing returns. The choice of $\gamma$ models the diminishing returns with the increase in layer decisions.

The choice of $\beta>1$ is motivated by multiple factors. The first is to help decrease optimal rate decisions for the different chunks. Suppose the rate decision for chunk $l$ is $r_l$ and that for chunk $l+1$ be $r_{l+1}$. For $\beta=1$, both $(r_l, r_{l+1}) = (1,2) $ and $(r_l, r_{l+1}) = (2,1) $ can be optimal decisions while for $\beta>1$, only $(r_l, r_{l+1}) = (1,2) $  could be optimal. The second is that the higher rates being pushed for the later chunks help in giving more time to avoid the bandwidth fluctuations due to incorrect bandwidth prediction. The third is that since the higher rates are pushed for later chunks (subject to buffer constraints), the quality variations would be reduced.

\begin{figure*}
\centering
\includegraphics[trim=0in 2in 0in 2in, clip,width=\textwidth]{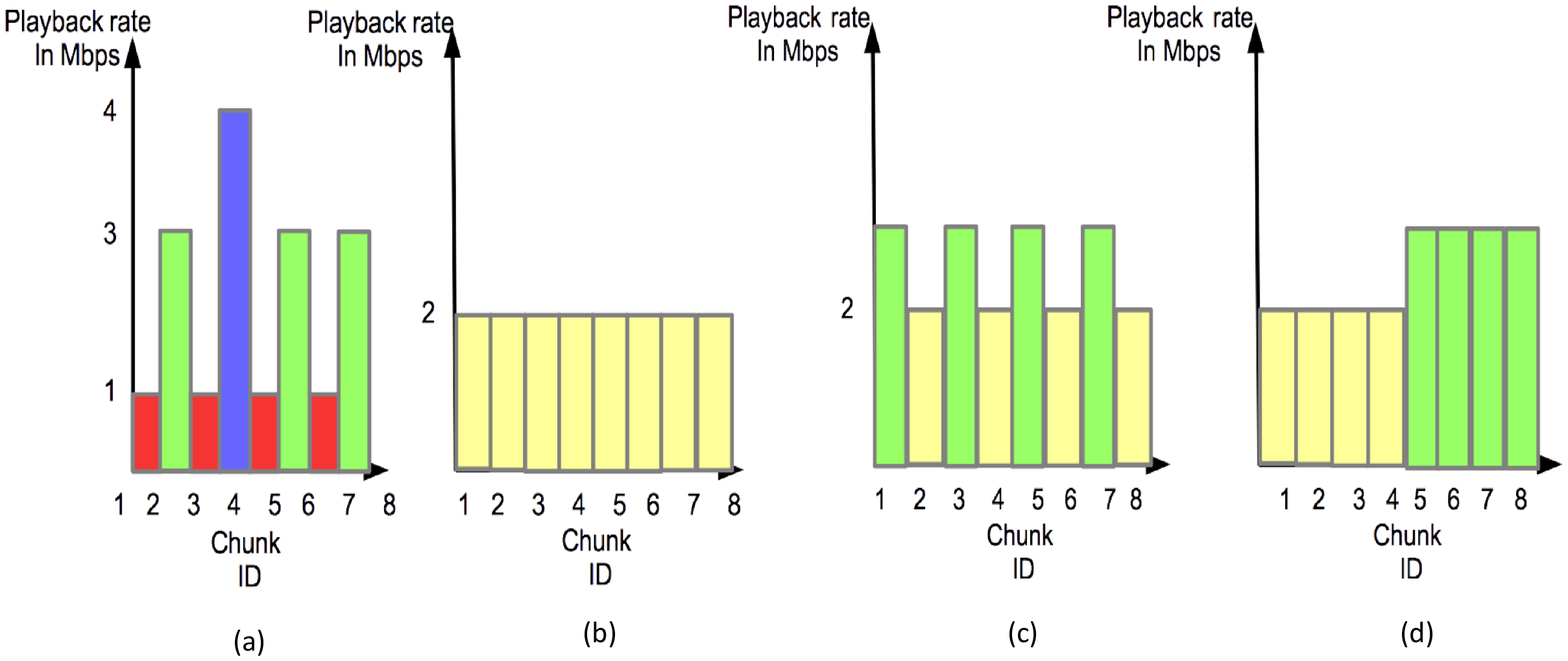}
\caption{(a) Playback Policy 1, (b) Playback Policy 2, (c) one of the optimal solution for the proposed objective function when $\beta=1$, and (d) the optimal solution to the proposed objective function when $\beta > 1$}
\label{fig:ex1}
\end{figure*}

In order to illustrate the intuition behind the above parameters ($\beta$ and $\gamma$), we present an example as shown in Fig.~\ref{fig:ex1}. In the example, we assume a video of 8 chunks (1s each), infinite buffer, and a bandwidth trace such that the two fetching Policy 1 and Policy 2, as shown in Fig.~\ref{fig:ex1}-(a) and Fig.~\ref{fig:ex1}-(b) are feasible. Summing up the playback qualities of the 8 chunks for the first policy will yield 17Mb. On the other, it will yield 16Mb for the second fetching policy. Therefore, if we are maximizing the average quality, then the first fetching policy is the better choice. However, since the user QoE is concave with the playback quality, it is beneficial to increase the lower rates since receiving chunks at the lowest quality is not preferable. To model the diminishing returns with increasing quality, the parameter $\gamma$ is chosen which would favor Policy 2 as compared to Policy 1. Further, we note that Policy 1 has high quality variations as compared to Policy 2. For example, switching from the $4^{th}$ to the $1^{st}$ layer quality in chunk 4 and 5 which  is an un-pleasing to the eye and decreases QoE. Therefore, pushing more chunks to the second layer quality level will be more preferable. The choice of $\gamma$ will prioritize fetching a lower layer of a chunk as compared to higher layer of another chunk, thus favoring Policy 2 to Policy 1. 


In order to describe the effect of $\beta$, we assume that there is a bandwidth trace such that the two fetching policies shown in Fig.~\ref{fig:ex1}-(c) and Fig.~\ref{fig:ex1}-(d) are feasible. Let Fig.~\ref{fig:ex1}-(c) shows one of the optimal solution when $\beta=1$, which may not longer be optimal for $\beta>1$. Let  Fig.~\ref{fig:ex1}-(d) shows the optimal solution when $\beta > 1$. We can clearly see that both solutions achieve the same average playback rate when $\beta=1$. Since both solutions are optimal for $\beta=1$, the choice of $\beta>1$ helps in providing unique optimal choice of chunk rates. Further, since the higher rates are allocated to later chunks, the prediction error can be handled better. For example, if the actual bandwidths for  first few time slots is lower, then the first chunk cannot be obtained at higher quality thus impacting decisions for policy in Fig. \ref{fig:ex1}-(c). Assuming unbiased error, the actual bandwidths for later time slots may be better, thus giving more time to still fetch the later chunks at higher quality for the policy in Fig. \ref{fig:ex1}-(d). Finally, we see that the first solution suffers from higher switching rate as compared to the second. In order to favor the second solution, we choose $\beta>1$, and with this choice we reduce the number of quality switches from $4$ to $1$. Even though we choose $\beta = 1+\epsilon$ for $\epsilon$ small, the proposed algorithm is also one of the optimal solutions for $\beta=1$.

\end{thisnote}

\section{Example for  Skip Based Streaming Algorithm}\label{skipex}

\begin{figure*}[htbp]
	\begin{minipage}{.28\textwidth}
		\includegraphics[trim=0.3in 0in 0.3in 0.2in, clip,width=\textwidth]{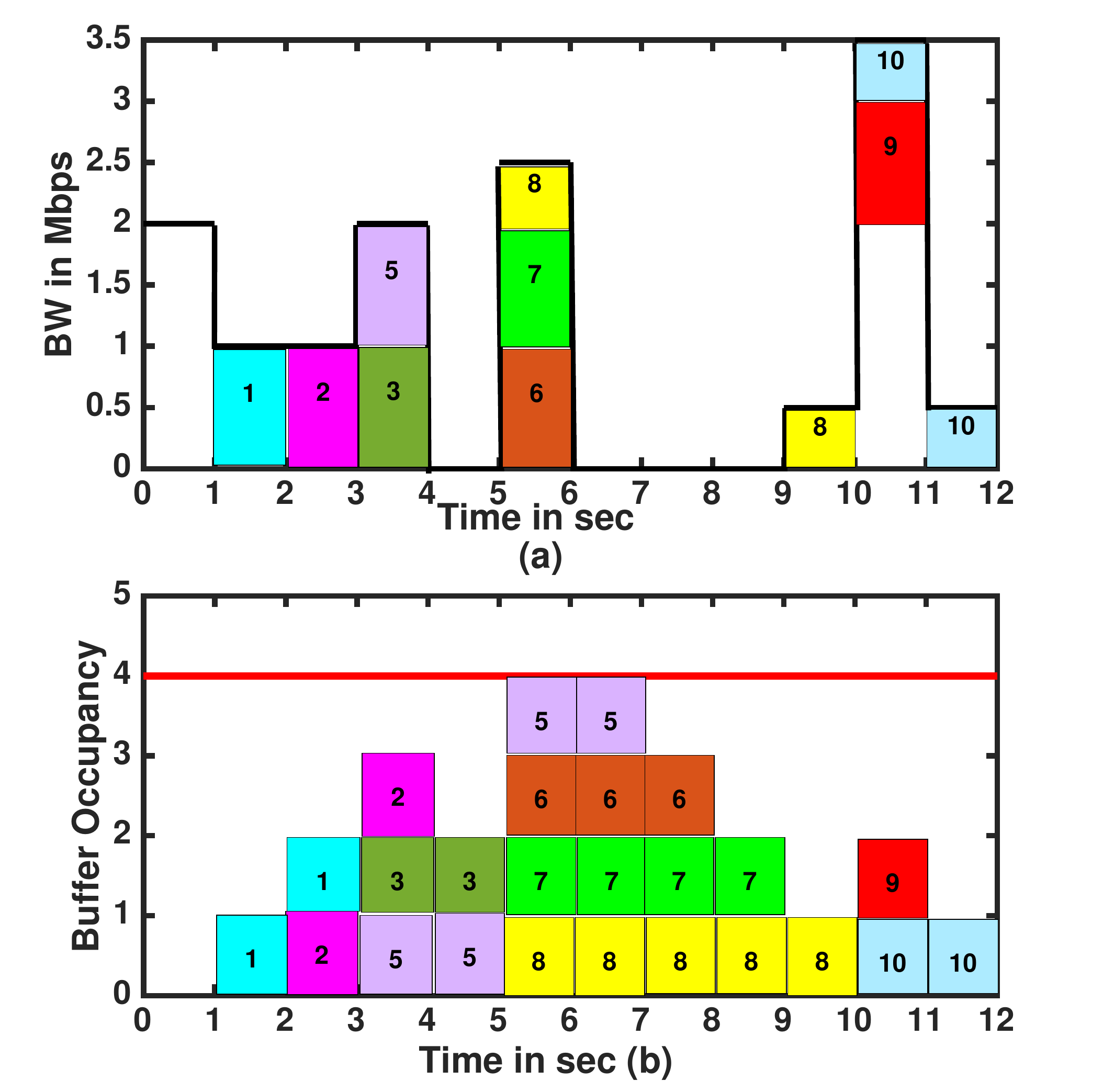}	
		\vspace{-.1in}
		\caption{Backward Algorithm for Base Layer, (a) bandwidth profile and bandwidth utilization, (b) Buffer occupancy }
		\label{fig:ex11}
	\end{minipage}
	\hspace{.1in}
	\begin{minipage}{.28\textwidth}
		\includegraphics[trim=0.3in 0in 0.3in 0.2in, clip,width=\textwidth]{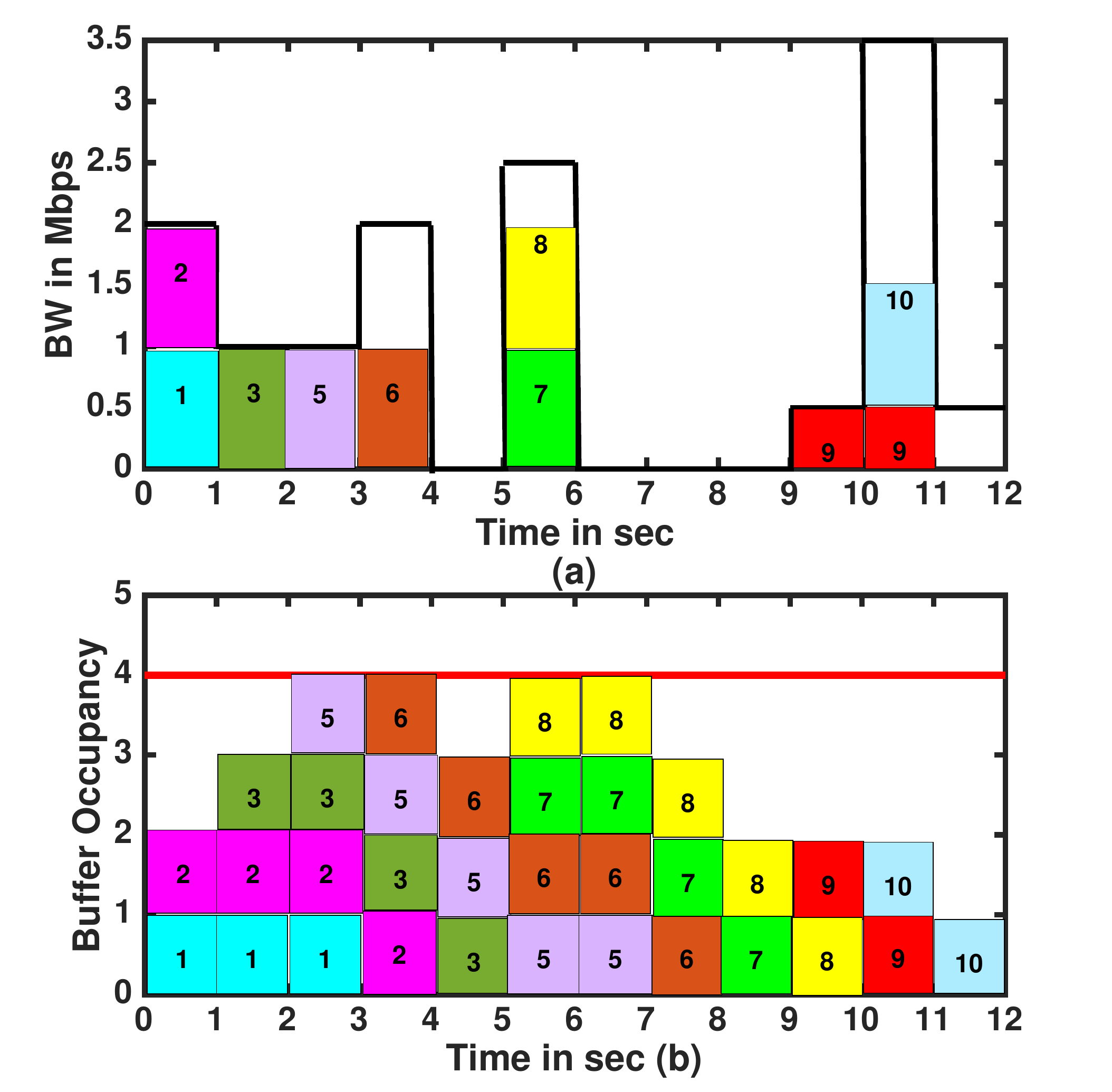}
		\vspace{-.1in}
		\caption{Forward Algorithm for E1, (a) bandwidth profile  and bandwidth  utilization, (b) Buffer occupancy}
		\label{fig:ex12}
	\end{minipage}
	\hspace{.1in}
	\begin{minipage}{.28\textwidth}
		\includegraphics[trim=0.3in 0in 0.3in 0.2in, clip,width=\textwidth]{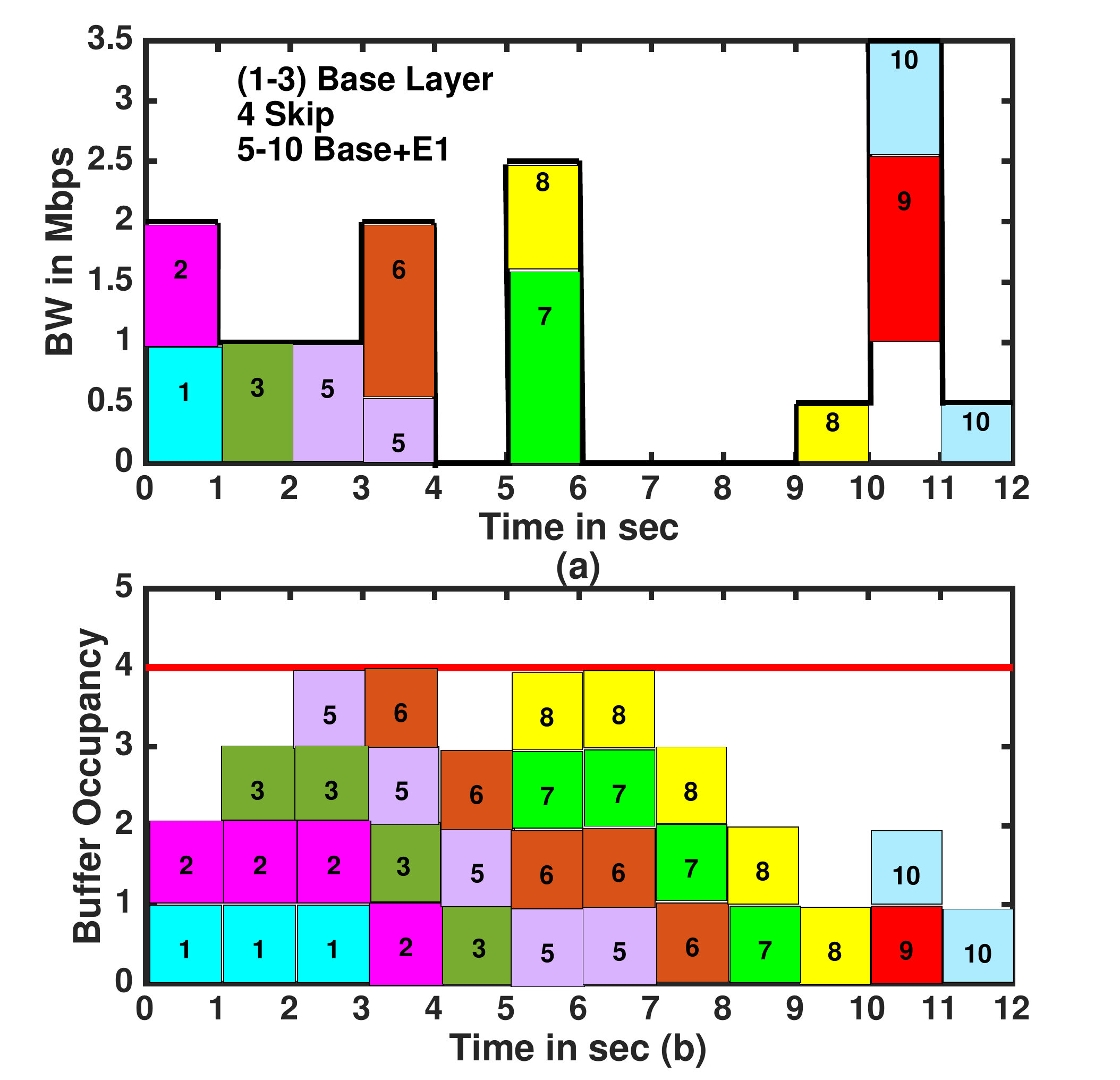}
		\vspace{-.1in}
		\caption{Backward Algorithm for E1 layer, (a) bandwidth profile  and bandwidth utilization, (b) Buffer occupancy}
		\label{fig:ex13}
	\end{minipage}
	
\end{figure*}

Assume we want to schedule 10 chunks, each with BL of size 1Mb, and E1 of size 0.5Mb. We assume that chunk duration is 1s, startup delay is 3s, and buffer length is 4s. The bandwidth profile, layer decisions, and buffer occupancy are shown in Fig. \ref{fig:ex11}-\ref{fig:ex13}. Fig. \ref{fig:ex11}(a-b) shows how backward algorithm works, where Fig \ref{fig:ex11}-a shows how bandwidth profile is utilized, and Fig \ref{fig:ex11}-b shows how the buffer occupancy changes with time. The algorithm simulates fetching chunks starting by fetching the last chunk (chunk 10) at its deadline time slot ($deadline(10)=(10-1)*1+3=12$). Chunks 10, 9, 8, 7, 6, and 5 were all fetched. However, if chunk 4 is fetched the buffer constraint at 6th time slot would be violated since chunks 4-8 need to be in the buffer at 6th time slot if all of them are fetched (see Fig.\ref{fig:ex11}-b). Therefore, this chunk is skipped. Finally, chunks 1-3 are also fetched since they do not violate any of the constraints. 
 The forward algorithm is then run in order to find the lower deadlines of the non-skipped chunks. As Fig. \ref{fig:ex12}(a-b) shows, forward algorithm simulates fetching non-skipped chunks in sequence (earlier deadline first) such that the bandwidth and the buffer constraints are not violated. For example, the lower deadline of chunk 10, $t(10)=11$, and of chunk 8 is 6, $t(8)=6$. Moreover the amount of data for chunks 7 and 8 downloaded at time 6 should not exceed $B(6)$; otherwise chunk 7 can't be fully downloaded.

Last step is shown at Fig \ref{fig:ex13}(a-b). In this step, the backward algorithm is run again. We start from chunk number $10$ but we consider the time slots from $t(10)$ to $deadline(10)$ ($11$th and $12$th time slots). If this time has enough bandwidth to fetch both BL and E1 of the $10$th chunk, then the final decision is to fetch chunk $10$ at E1 quality. According to this example, final decision is to fetch chunks 1-3 at BL, skip chunk 4, and chunks 5-10 are fetched at E1 quality. 

\section{Proof of Lemma \ref{lem:skip:beta1}}\label{apdx:skip:beta1}

We first note that using the backward algorithm, $n$th layer of chunk $i_s$ is skipped in three scenarios, which are described as follows.

\noindent {\bf Case 1: } When the $(n-1)$th layer of the chunk $i_s$ is not fetched. In this case no other algorithm can fetch the $n$th layer of chunk $i_s$ due to constraint (\ref{equ:c2eq11}) violation.

\noindent {\bf Case 2: } When the residual bandwidth in $(t(i_s)$ to $deadline(i_s))$, ignoring the bandwidth used to fetch up to the ($n-1$)th layer of chunks less than $i_s$ (lower layer decisions) and the bandwidth used to fetch up to the $n$th layer of chunks greater than $i_s$ (current layer decisions for later chunks) is not enough to fetch the $n$th layer of chunk $i_s$. Mathematically, let's define $\pi_{prior}$ as what is reserved to fetch up to the $(n-1)$th layer of chunks $< i_s$, and $\pi_{later}$ as what is reserved to fetch up to the $n$th layer of chunks $> i_s$. Since decision of the current layer size starts from back, then the $n$th layer of the chunk $i_s$ is skipped when: 

\begin{equation}
\sum_{j=t(i_s)}^{deadline(i_s)}B(j) - \pi_{prior} -\pi_{later} < X_{n-1}+Y_{n}
 \end{equation}

\noindent {\bf Case 3: } When $n$th layer of chunk $i_s$ is obtained, the buffer constraint gets violated. {\em i.e.}, when chunks $i_s$ to $i_e$ (excluding the skipped chunks in between) need to be in the buffer at a certain time in order not to skip $n$th layer of $i_s$, but their length is longer than the buffer duration. Mathematically there is a segment of chunks, $I=\{i_s,.....,i_e\}$  with length $N=i_e-i_s+1- \sum_{i_s}^{i_e} {\bf1}\{ Z_0(i)=0\} $, such that:
 \begin{equation}
N \cdot L > B_m
 \end{equation}
 
 We note that the forward algorithm fetches chunks at the earliest possible time while the backward algorithm fetches chunks at the latest possible time. We will now show that the backward algorithm has no larger skips at any layer (given previous layer quality decisions) as compared to any other feasible algorithm. 
 
  
 
Any  skips which are due to Case 1 will be the same for any feasible algorithm. Thus, we do not consider skips that are of Case 1.  For the skip of Case 2 and Case 3, we first consider the last skip by our algorithm. We note that since the backward algorithm will consider fetching  $n$th layer of any chunk if the cumulative bandwidth up to the deadline of that chunk is higher than the $n$-th layer size ($Y_n$), then, if there is an $n$th layer skip at $i_s$ (the cumulative bandwidth up to the deadline of $i_s$ is less than the $n$th layer size), any algorithm will have a skip for a chunk $i\ge i_s$. In other words, if another algorithm fetches the $n$th layer of chunk $i_s$, it must have skipped the $n$th layer of a chunk with an index $i$ such that $i > i_s$. Further note that skipping the $n$th layer of chunk $i_s$ as compared to chunk $i$ will allow for more bandwidth up to the deadline of the earlier chunks.  Thus, if there is another skip using the backward algorithm, there must be another skip at or after the time of that skip in any other algorithm. Thus, we see that the number of skips in any feasible algorithm can be no less than that in the proposed algorithm.

\section{Proof of Lemma \ref{lem:skip:betage1}}\label{apdx:skip:betage1}

We note from Appendix \ref{apdx:skip:beta1} that for any skip of backward algorithm, there must be a skip or more for any other feasible algorithm. Further, since any feasible algorithm that has more number of skips can't achieve higher objective, we consider algorithms with a number of skips that is equal to what is achieved by the backward algorithm.

 In this case, we note from Appendix \ref{apdx:skip:beta1} that if the ordered set of skips for the backward algorithm are $i_1, i_2, \cdots, i_H$  and for any other feasible algorithm are $j_1, j_2, \cdots, j_H$,  then $i_k\le j_k$ for all $k=1, \cdots, H$. Using this,  and knowing that $\beta \geq 1$, we see that the result in the statement of the Lemma holds.

\section{Proof of Theorem \ref{theorem: theorem1}}\label{apdx:them1}
We note that the conditions, $0<\gamma < 1$ and  (\ref{basic_gamma_1}) depict the  priority of horizontal scanning over vertical scanning. It tells that fetching the base layer of the first chunk which has the lowest priority among all chunks achieves higher objective than fetching all the higher layers of all the chunks. Therefore, if $\gamma$ and $\beta$ satisfy the stated conditions, the sequential scanning among different layers starting from the base layer up to the highest enhancement layer can show the optimality of the proposed algorithm. 


{Lemma~\ref{lem:skip:betage1} proves that running the backward algorithm for the $n$th layer given lower and upper deadlines of every chunk achieves the optimal objective for the $n$th layer. Applying this to the base layer will give the optimal number of skips.}

{According to proposition \ref{noSkipPro}, running the forward algorithm for sizes decided by the backward algorithm on base layers will fetch chunks at their earliest. Therefore, the lower deadline of every chunk such that all chunks are fetched according to the base layer is found by the forward algorithm.}

{Given the optimal lower deadlines of all chunks, running the backward algorithm on $E1$ layer would produce the optimal $E1$ decision without violating lower layer decisions.  Scanning sequentially up to $M$th layer would yield the optimal decisions of layers $0$ to $M$, and that concludes the proof.}

\section{No-Skip Layered Bin Packing Adaptive Algorithm}
\label{sec: noSkipCode}

The pseudo-code for the no-skip Layered Bin Packing Adaptive Algorithm is described in Algorithm \ref{algo:basenoskip}, which uses Algorithms \ref{algo:basefwdnoskip} and \ref{algo:basebacknoskip}. The algorithm works as follows. First, Base Layer Forward algorithm is run to find the deadline of the last chunk (chunk $C$). Then, Base Layer Backward algorithm is run with the above deadline of last chunk to determine the number of stalls before each chunk.  Forward algorithm is run after that to simulate fetching chunks in sequence as early as possible, so the lower deadline of every chunk is found. Finally, backward and forward algorithm of skip version is run the same way we described for the skip version algorithm per layer since skips are allowed for enhancement layers.

\begin{figure}[h]
		\begin{minipage}{\linewidth}
			\begin{algorithm}[H]
				\small
				\begin{algorithmic}[1]
					
    \STATE $d(i)=0, \forall i$
   \STATE $deadline(i)=(i-1)L+s+d(i)$
    \STATE {\bf Base Layer decision:}
    \STATE $[d,x] = BaseLayer forward Algorithm(B, X, C,d, deadline,$ $B_m, bf)$
    \STATE $deadline(i)=(i-1)L+s+d(C)$
     \STATE $d_f=Base Layer Backward Algorithm(B, X_0, x, C, d,$ $deadline, B_m, bf)$
      \STATE $deadline(i)=(i-1)L+s+d_f(i)$
      \STATE $[t,x,e] = forwardAlgo(B, X, C, deadline, Bm,bf, I_0)$
  \STATE {\bf Enhancement layer decisions:}
   \STATE {\bf For every enhancement layer $n$}
  \STATE $[X, I_n]= backward Algorithm(B, X,X_{n}, C, L, deadline,$ $B_m,bf,t,c,x,e)$
    \STATE $[t,a,e] = forward Algorithm(B, X, C, deadline, Bm,bf, I_0)$
   				\end{algorithmic}
				\caption{No-Skip Layered Bin Packing Adaptive Algorithm }\label{algo:basenoskip}
			\end{algorithm}
		\end{minipage}
		\vspace{-.2in}
	\end{figure}
	
	\begin{figure}[hbtp]
		\begin{minipage}{\linewidth}
			\begin{algorithm}[H]
				\small
				\begin{algorithmic}[1]
				\STATE {\bf parameters:} $bf(j)$: buffer length at time $j$, $d(i)$: stall time of chunk $i$.
				\STATE {\bf Output:} $d(i)$: stall time of chunk $i$.

   \STATE $ j=1$
 \STATE $i=1$
    \WHILE {chunk $C$ is not fetched yet}
      		\IF{$(i > 1$ and $d(i) < d(i-1))$}
			\STATE $d(i)=d(i-1)$
			\STATE $deadline(i)=(i-1)L+s+d(i)$
		\ENDIF
            \IF{$(bf(j) = Bm)$}
                    \STATE $j=j+1$
                    \STATE continue
                \ENDIF

                 \STATE $fetched=min(B(j), X(i))$

                 \STATE $B(j)=B(j)-fetched$
                 \STATE $X(i)=X(i)-fetched$
                 \IF{$X(i) > 0$}
                 	 \STATE $bf(j)=bf(j)+L$
	 	\ENDIF
                 \IF{$X(i)=0$ and $j <= deadline(i)$}
                      \STATE i=i+1
                \ELSIF{$X(i)=0$ and $j >deadline(i) $}
                \STATE $d(i)=d(i)+j-deadline(i)$
                      \STATE $i=i+1$
                \ENDIF

                 \IF{$B(j)=0$}
                    \STATE $j=j+1$\\
                \ENDIF

                      \ENDWHILE

				\end{algorithmic}
				\caption{Base Layer Forward Algorithm For No-Skip Streaming}\label{algo:basefwdnoskip}
			\end{algorithm}
		\end{minipage}
		\vspace{-.2in}
	\end{figure}

	\begin{figure}[hbtp]
		\begin{minipage}{\linewidth}
			\begin{algorithm}[H]
				\small
				\begin{algorithmic}[1]
				\STATE {\bf parameters:} $bf(j)$: buffer length in chunks at time $j$, $d(i)$: initial deadline of chunk $i$ which was found from running forward algorithm.
					\STATE {\bf Output:} $d_f(i)$: final stall duration of chunk $i$.
					
    \STATE {\bf Initilization:}
    \STATE $i=C$
   \STATE $ j=deadline(C)$

     \WHILE {$j > 0$ and $i > 0$}
     \IF{$(i < C)$}
     	\STATE $d_f(i)=d(i)-(d(i-1)-d_f(i-1))$
	\STATE $deadline(i)=(i-1)L+s+d_f(i)$
     \ENDIF
        \IF{$j <= deadline(i)$}
        \IF{($bf(deadline(i)) = B_m)$}
               \STATE $d_f(i)=d(i)-1$
               \STATE continue
          \ENDIF



               \STATE $fetched=min(B(j), X_0(i))$
                \STATE $X_0(i)=X_0(i)+fetched$
                \STATE $B(j)=B(j)-fetched$
                 \IF{$X(i) > 0$}
                 	 \STATE $bf(j)=bf(j)+L$
	 	\ENDIF

                \IF{$(X_0(i) = 0)$}
                    \STATE $i=i-1$
                \ENDIF

                \IF{$(B(j) = 0)$}
                    \STATE $j=j-1$
                \ENDIF

                   \ELSE
           \STATE $j=j-1$
        \ENDIF
\ENDWHILE

				\end{algorithmic}
				\caption{Base Layer Backward Algorithm For No-Skip Streaming }\label{algo:basebacknoskip}
			\end{algorithm}
		\end{minipage}
		\vspace{-.2in}
	\end{figure}

\section{Example for No Skip Based Streaming Algorithm}\label{noskipex}

\begin{figure*}[htbp]
	\begin{minipage}{.38\textwidth}
		\includegraphics[trim=0in 0in 0in 0in, clip,width=\textwidth]{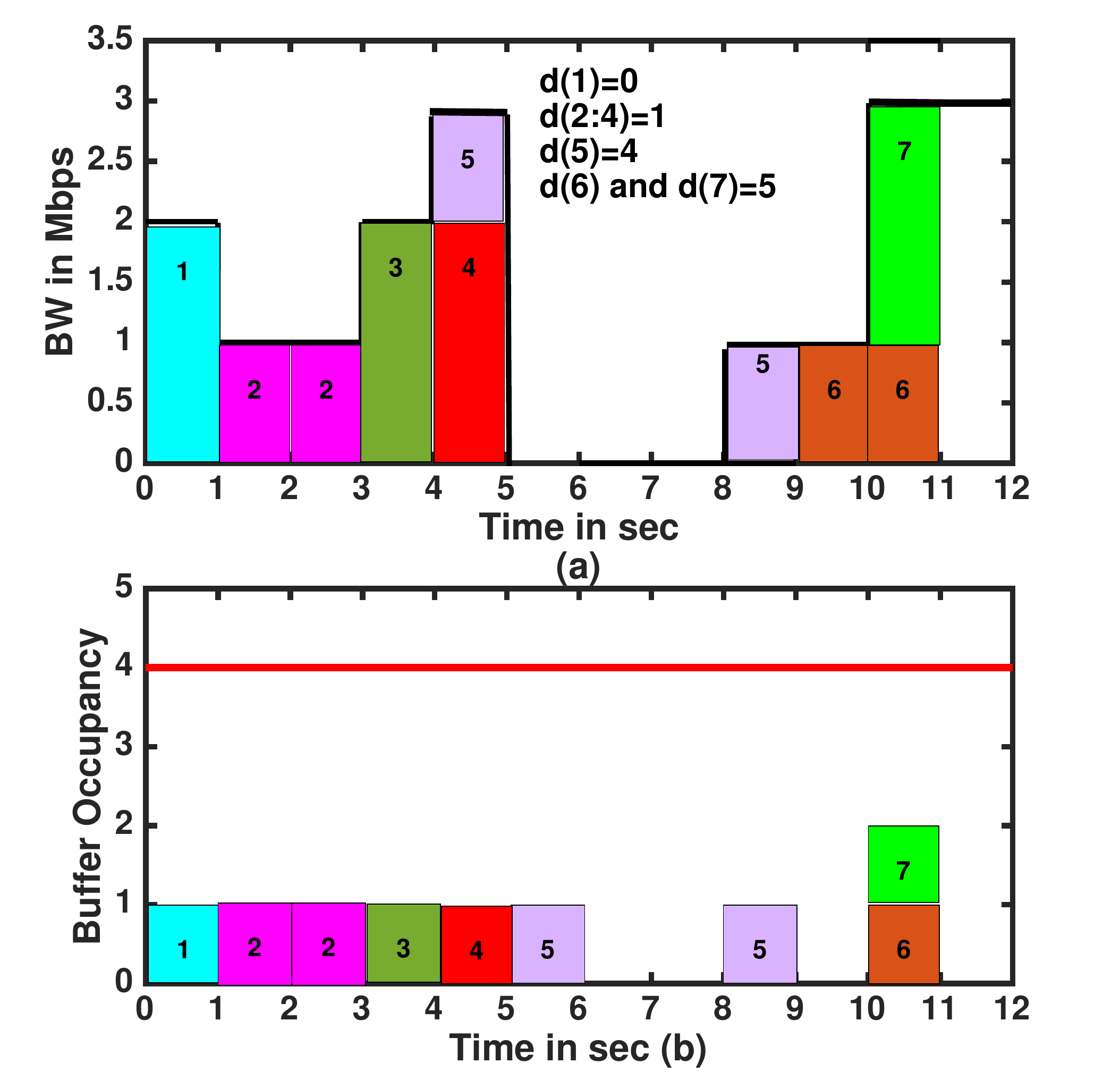}	
		\label{fig: ex21}
		\vspace{-.1in}
		\caption{Base Layer Forward Algorithm, (a) bandwidth profile and utilization, (b) Buffer occupancy }
		\label{fig:noskipex1}
	\end{minipage}
	\hspace{.1in}
	\begin{minipage}{.58\textwidth}
		\includegraphics[trim=0in 0in 0in 0in, clip,width=\textwidth]{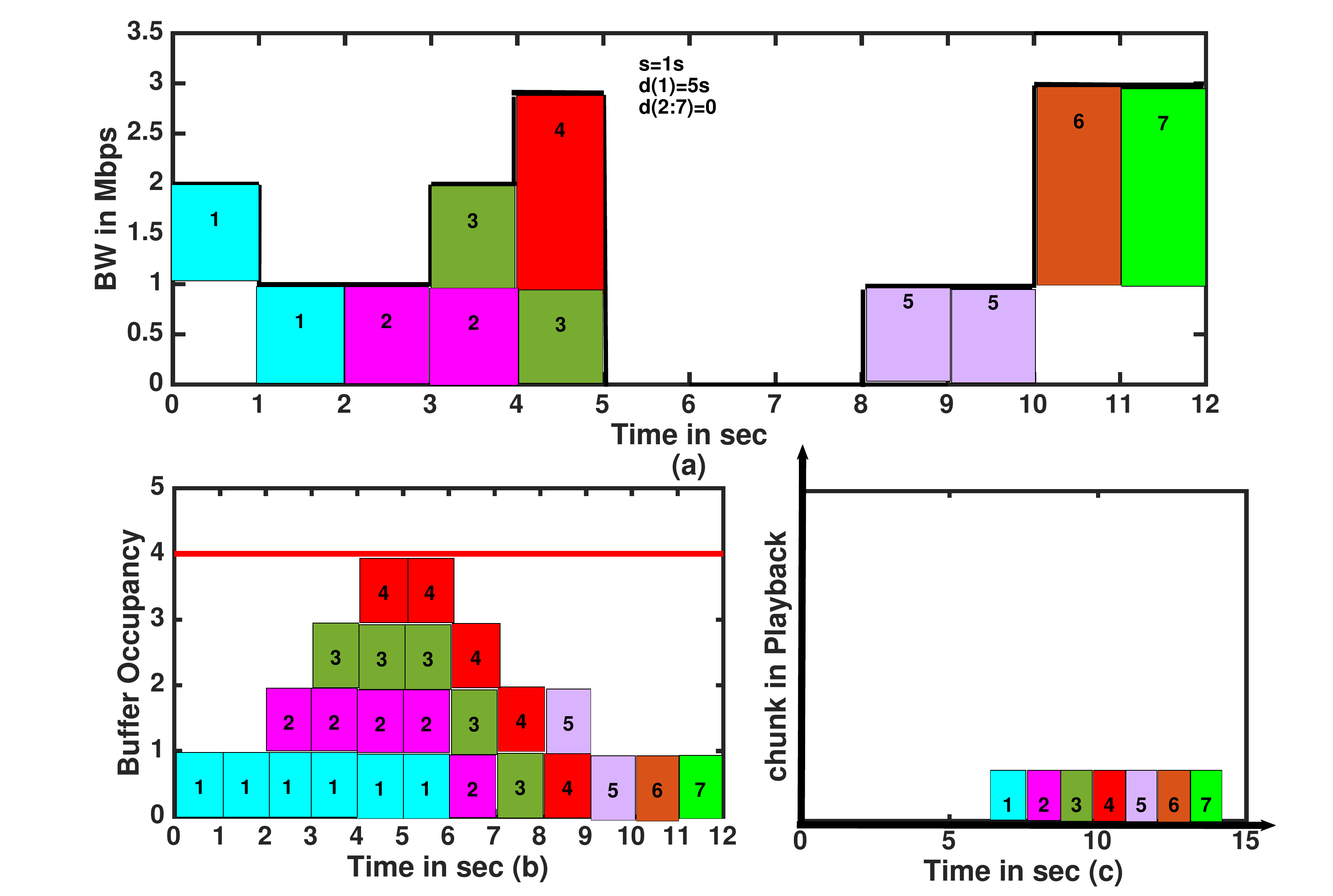}
		\label{fig: ex12}
		\vspace{-.1in} 
		\caption{Base Layer Backward Algorithm, (a) bandwidth profile and utilization, (b) Buffer occupancy, (c) playback time of chunks}
		\label{fig:noskipex2}
	\end{minipage}
\end{figure*}

Assume we want to schedule 7 chunks, each with BL of size 2Mb. We assume that chunk duration is 1s, initial startup delay is $s=1s$, so the initial deadline of every chunk is: $deadline(i)=(i-1)L+s$, and buffer length is 4s. The bandwidth profile, layer decisions, and buffer occupancy are shown in Fig. \ref{fig:noskipex1}-\ref{fig:noskipex2}. Fig. \ref{fig:noskipex1}(a-b) shows how base layer forward algorithm works, where Fig. \ref{fig:noskipex1}-a shows how bandwidth profile is utilized, and Fig. \ref{fig:noskipex1}-b shows how the buffer occupancy changes with time. The algorithm simulates fetching chunks in sequence starting from the first chunk. It chooses the time when a chunk is fully downloaded as the deadline of that chunk if fetching it exceeds its initial deadline (this is not final decision), so according to our example, $deadline(1)=1$, and $deadline(7)=11$  ({\em i.e.}, $d(1)=0$, and $d(7)=5$). Therefore, the stall time of the first chunk is zero $(d(1)=0)$, and the stall time of the $7$th chunk is 5s $(d(7)=5s)$. 

We next run the Base Layer Backward Algorithm to move the stalls to earliest possible times so that  all chunks can have more time to get their higher layers without violating $deadline(C)$. 
Fig. \ref{fig:noskipex2}-a shows how base layer backward algorithm starts fetching chunks in reverse order with $7$th chunk being the first one to fetch and starting point to fetch it is its deadline time slot ($deadline(7)$). All chunks can be fetched one after the other without stalls in between. We clearly see from playback figure (Fig. \ref{fig:noskipex2}-c) that all stalls are brought to the very beginning, so $d(1)=5$, and new startup delay is $s+d(1)=1+5=6s$. If we want to consider E1, we need to run forward algorithm at the base layer (not shown in the figure) to fetch chunks at their earliest and provide E1 backward algorithm with $t(i)$ and the base layer decisions. The rest is similar to the no skip algorithm since skips are allowed for higher layers. We would like to mention  that in this example we were able to bring all stalls to the beginning, but that is not always the case. In fact, if the buffer size of this example was $3s$ we would have a stall between the first and second chunk. 

\section{Proof of Lemma \ref{lemma: noSkipLemma1}}\label{apdx:lemma1Noskip}

We note that for any feasible algorithm, we can convert to an in-order feasible streaming algorithm with the same time of receiving the last chunk. For an in-order streaming algorithm, the proposed base layer forward algorithm is a greedy algorithm that only adds to the deadline if the chunk cannot be downloaded in time due to not enough bandwidth. Further, the algorithm fetches all the chunks at only base layer quality which would have the lowest possible fetching times. Due to this greedy nature, no other algorithm can fetch all the chunks before those obtained by this greedy algorithm. 

\section{Proof of Theorem \ref{thm:noskip}}\label{apdx:lthem2Proof}

{From Lemma~\ref{lemma: noSkipLemma1}, we see that No-Skip forward algorithm finds the minimum stall duration such that all chunks can be fetched at least at base layer quality. However, after finding the minimum stall duration, the remaining part of the algorithm is similar to the skip based streaming problem since skips are allowed for enhancement layers. The No-Skip based streaming is a special case of the skip based streaming in which the chunk deadlines are chosen such that there are no base layers skips. Therefore, the rest of the proof follows the same lines as the proof of Theorem~\ref{theorem: theorem1} in Section~\ref{skipalgo}.}
\section{Impact of Algorithm Parameters}
\label{sec:eval_para}

\begin{figure}
\includegraphics[trim=0.3in 0.1in 0.5in 0.3in, clip,width=0.48\textwidth]{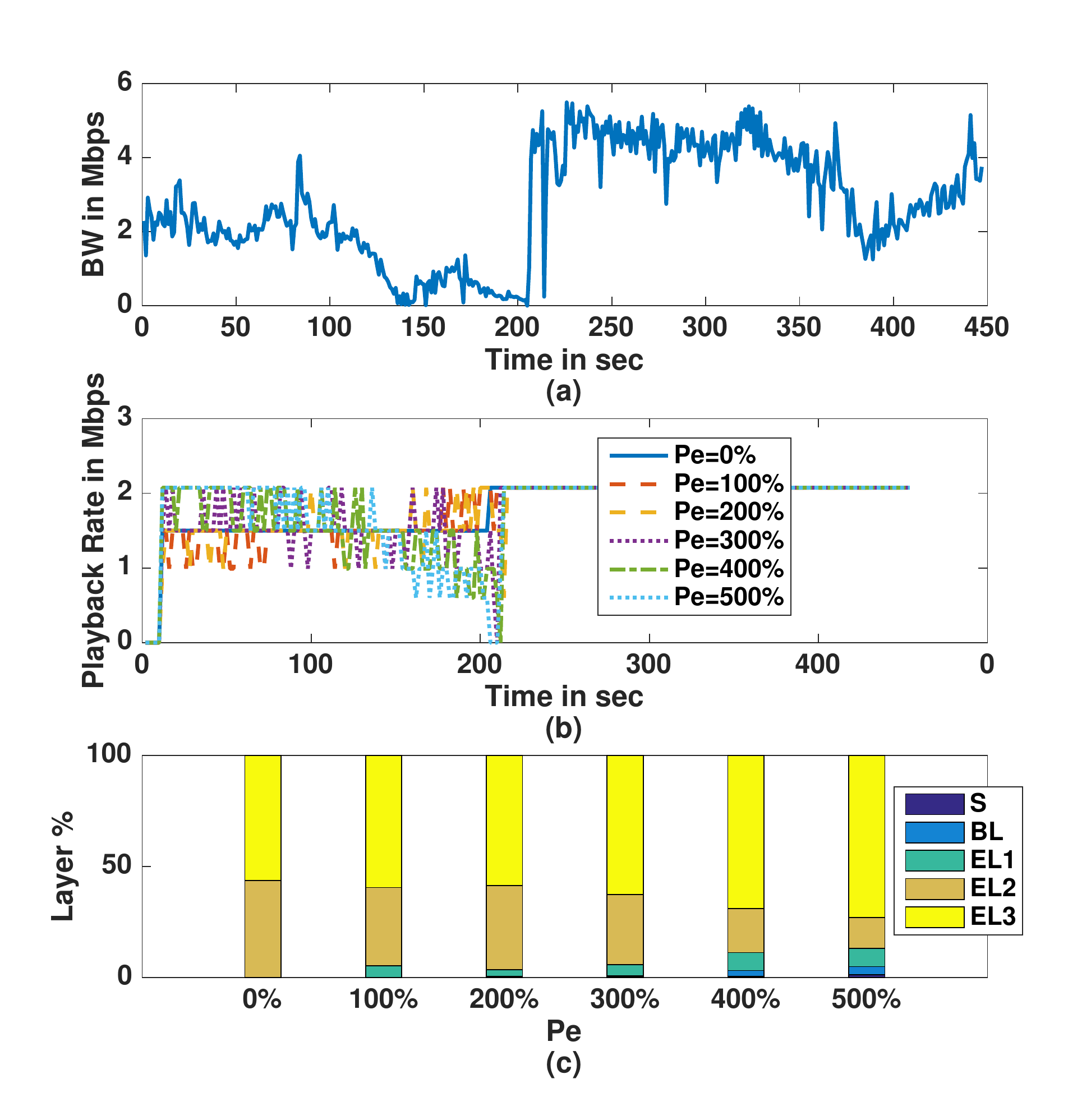}	
 \caption{Impact of bandwidth prediction accuracy: (a) bandwidth profile, (b) playback bitrate time series, and (c) layer breakdown (percentage of time of playback at different rates).}
 \label{fig : peCdf}
\vspace{-.1in}
\end{figure}

We now systematically study the impact of the key parameters (\eg the prediction window size and playout buffer size) as well as the bandwidth prediction accuracy on the algorithm performance.
As a case study, all experiments here use one representative trace (shown in Fig.~\ref{fig : peCdf}-a) chosen from the 50 traces because of its
highly variable network bandwidth.
This allows us to investigate the detailed time series of how our algorithm behaves over time under challenging network conditions. Using other traces with highly variable bandwidth yields qualitatively similar conclusions. We consider skip-based streaming in this case study.

We assume the video is CBR encoded, so that we can use the nominal rate of every layer as the layer size of every chunk. The reason we assume CBR per layer in this section is that we plot the playback, so we need to see the changes of quality that are corresponding to prediction error, short prediction, and short buffer size instead of the changes related to different chunk sizes.

{\bf Impact of bandwidth prediction accuracy.}
Our algorithm leverages network bandwidth prediction whose accuracy affects the streaming quality.
%
%
To study the impact of the prediction accuracy, we varied the prediction error $pe$ (introduced in~\S\ref{sec:eval_skip}) from 0\% to 500\% with a step of 100\% (Note that we threshold the negative bandwidths obtained with more than 100\% prediction error to zero.).
%
%
We assumed a 5-second startup delay, the chunk size is 2 second (CS=2s), and 2-minute buffer size. 
Fig.~\ref{fig : peCdf}-b shows the playback time series for different $pe$ values. The X axis is the timeline and the Y axis is the chunk playback bitrate where each discrete level cumulatively represents a layer (the lowest value for BL, the second lowest value for EL1, \etc).
%
%
%
%
%
Fig.~\ref{fig : peCdf}-c shows the breakdown of the number of chunks fetched at different layers.
As prediction error increases, skips and the number of chunks fetched at only the base layer increase. Further, the number of chunks at the highest layer also increase efficiently utilizing the total bandwidth. 
As we see from Fig.~\ref{fig : peCdf}-b, the  chunks that drop quality with increasing error correspond to the first half of the video when the bandwidth is low. Overall, Fig.~\ref{fig : peCdf}-b and Fig.~\ref{fig : peCdf}-c indicate that the impact of prediction error on the chunk layer distribution is small.

{\bf Impact of prediction window size $W$.}
We also use the  bandwidth trace in Fig.~\ref{fig : peCdf}-a to
demonstrate the impact of $W$.
To better reflect realistic scenarios,
we varied both the window size $W$ and the prediction error $pe$ by assuming
that a larger $W$ has a higher $pe$.
We choose $(W,pe) \in \{(10,25\%), (20,50\%), (60,100\%), (120,200\%)\}$.
%
We also evaluate the optimal scheme where $W=\infty$ and $pe=0$.

\begin{figure}
\includegraphics[trim=0.3in 0.1in 0.5in 0.3in, clip, width=0.48\textwidth]{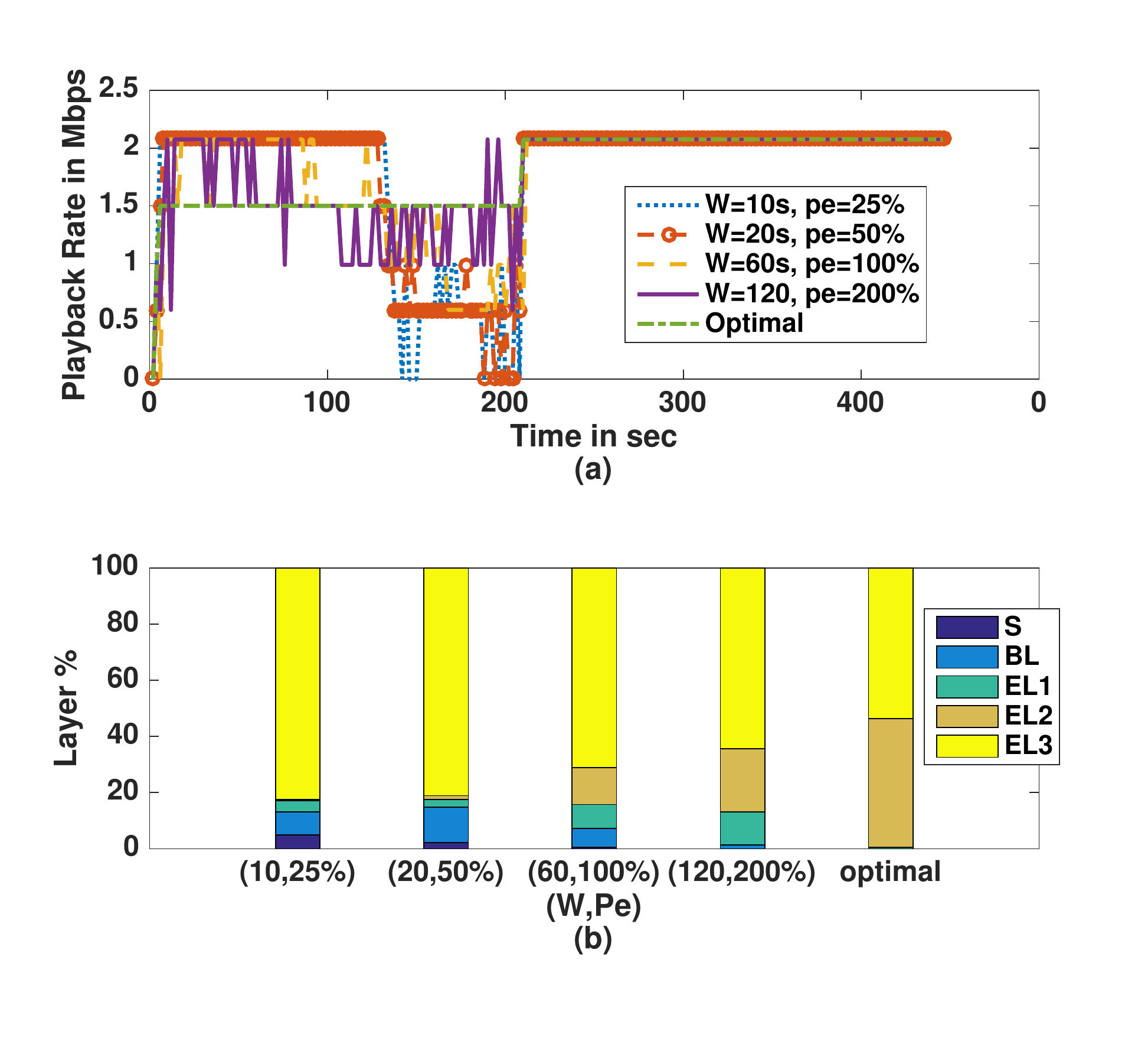}	
  \vspace{-.3in}
 \caption{Impact of the prediction window $W$: (a) playback bitrate time series, and (b) layer breakdown.}
  \label{fig : wPlot}
  \end{figure}

The results are plotted in Fig.~\ref{fig : wPlot}.
%
%
With a small $W$, it is difficult for our algorithm to leverage enough future bandwidth information.
%
%
As a result, the client aggressively fetches chunks at EL3 in the beginning but later on (during the period from 150 to 200s) suffers from lower quality and even skips due to bandwidth starvation, as shown in Fig.~\ref{fig : wPlot}-a.
This issue is mitigated by introducing buffer threshold, such that if the buffer occupancy is less than this threshold, quality decisions are reduced by 1 layer, and increasing $W$ so that the player has more visibility of future network conditions.
%
Fig.~\ref{fig : wPlot}-b shows that by increasing $W$ (even with a higher $pe$), we are getting closer to the offline scheduling results in term of the playback quality and the number of skips experienced.

\begin{figure}
\includegraphics[trim=0.3in 0.1in 0.5in 0.6in, clip,width=0.48\textwidth]{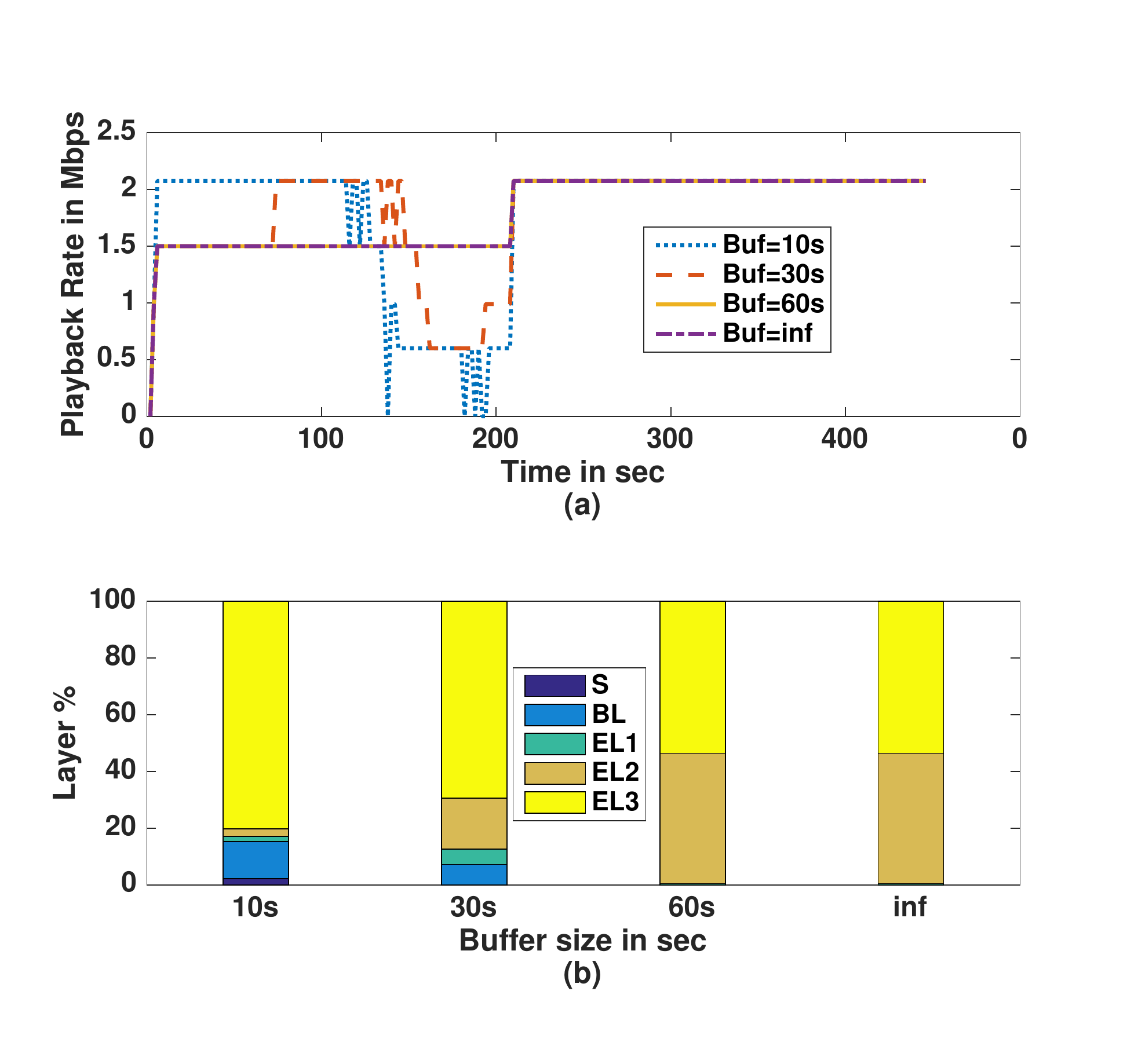}	
 \caption{Impact of the playout buffer size: (a) playback bitrate time series, and (b) layer breakdown.}
 \label{fig : bufCdf}
\end{figure}

{\bf Impact of the playout buffer size.}
%
The client-side playback buffer size affects the streaming performance:
a small buffer may limit the amount of the video contents that can be fetched;
while a large buffer causes bandwidth waste when the user abandons the video.
%
We again use the same bandwidth profile in Fig.~\ref{fig : peCdf}-a to
study the impact of the buffer size ($pe=0$, $W=\infty$, chunk size of 1s).
%
%
%
We consider buffer sizes of 10s, 30s, 60s, and $\infty$. 
%
Fig.~\ref{fig : bufCdf} shows that the buffer size does play a role in determining the performance in particular when the buffer is small.
Having a small buffer size of 10s causes a larger fraction of chunks to be delivered at EL3,
at the cost of more base layer chunks and skips, compared to scenarios with larger buffer sizes.
This is because shrinking the playout buffer essentially reduces the player's ``visibility'' of the distant future, leading to an effect similar to that caused by reducing $W$.
As the buffer size increases, we see fewer skips, few layer switches, and more chunks being delivered at higher layers (EL2 and above).
%
%


\begin{thisnote}
{\bf Impact of the buffer threshold $B_{min}$.} The parameter $B_{min}$ controls being overly optimistic when the playback buffer has less number of chunks. When the buffer is less than $B_{min}$, we drop the highest layer that was decided to be fetched (unless the decision is fetching only the base layer) using the proposed optimization problem.  We now see the  impact of $B_{min}$ on the percentage of chunks that are skipped. Table \ref{tab : bmin} illustrates the percentage of skipped chunks and the average playback rate for $(W, P_e, B_{max}) = (10,25,20)$ and $(20,50,20)$. For both these parameter choices, we evaluate the scenarios when $B_{min}$ is 0, 5, or 10s. We note that the number of skipped chunks improve with increasing $B_{min}$. However, there is a diminishing return. A larger $B_{min}$ also penalizes on the rates at which the chunks are fetched. Since the reduction of the stall duration is a bigger priority, the choice of $B_{min}$ should not go beyond that needed to achieve the lowest skip duration. Since this requires trial and error for each profile, we chose $B_{min}=B_{max}/2$ in our evaluations.

\begin{table}[htb]
	  \vspace{-.1in}
  \centering
  \caption{Impact of the buffer threshold $B_{min}$}
  \begin{tabular}{|c|} \hline
  $W=10$, $Pe=25$, $B_{max}=20$\\ \hline
  \end{tabular}
  \begin{tabular}{|c|ccc|} \hline
   $B_{min}$ in seconds & 0 & 5 & 10 \\ \hline
   Percentage of Skip Duration &  8.02& 7.47& 7 \\ \hline
   Average Playback Rate(Mbps)& 1.36 & 1.34& 1.31\\ \hline
    \end{tabular}
     \begin{tabular}{|c|} \hline
   $W=20$, $Pe=50$, $B_{max}=20$\\ \hline
    \end{tabular}
    \begin{tabular}{|c|ccc|} \hline
   $B_{min}$ in seconds & 0 & 5 & 10 \\ \hline
   Percentage of Skip Duration &  6.449& 6.13& 5.7 \\ \hline
   Average Playback Rate(Mbps)& 1.36 & 1.34& 1.31\\ \hline
  \end{tabular}
  \label{tab : bmin}
  \vspace{-.1in}
\end{table}

\end{thisnote}

\begin{figure}
	\vspace{-.1in}
	\includegraphics[trim=0.1in 0in 0.5in 0.1in, clip,width=0.48\textwidth, height=1.7in]{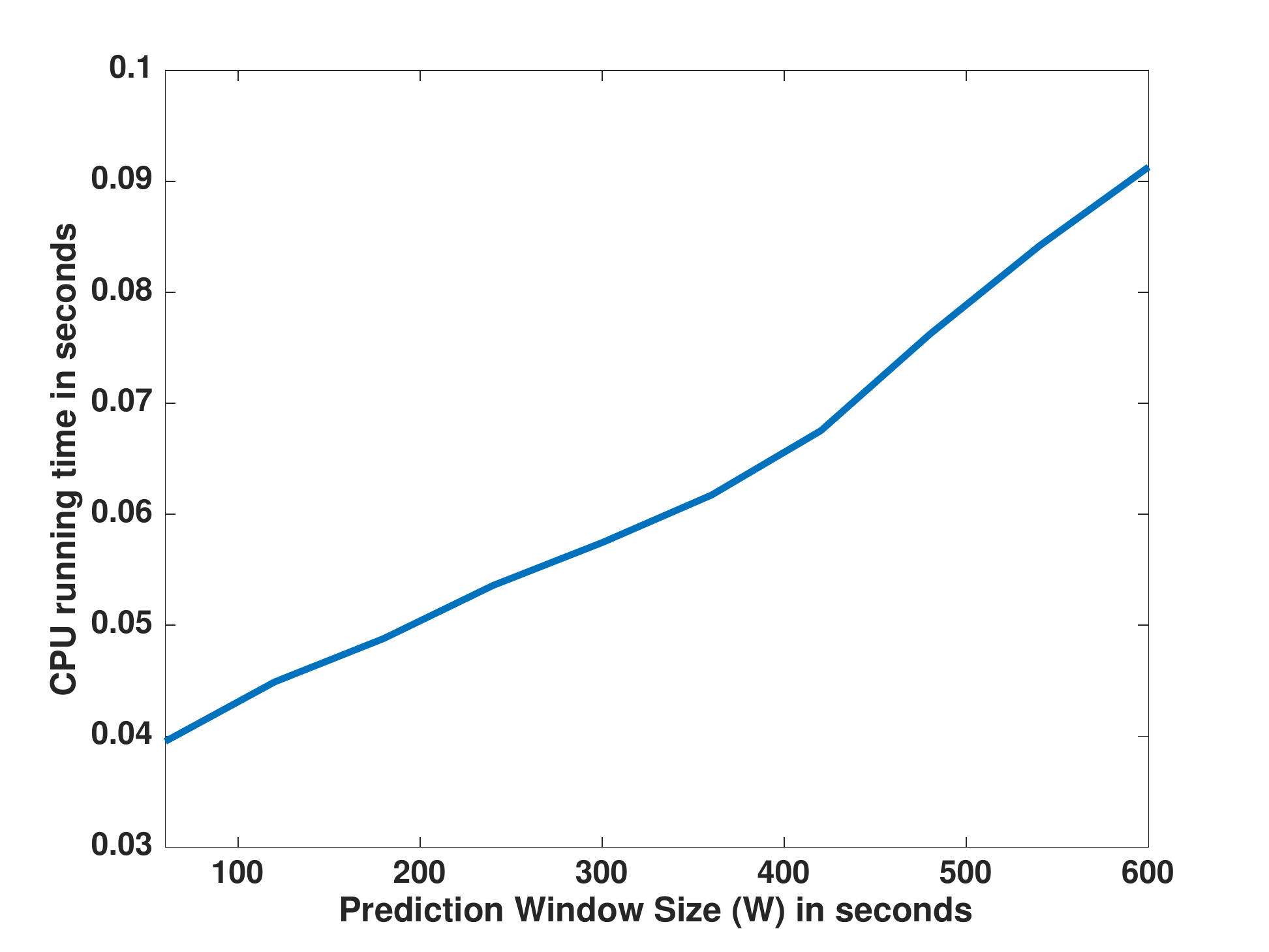}	
	\caption{CPU time with respect to the prediction window}
	\label{fig : cpuTime}
	\vspace{-.1in}
\end{figure}

\section{Computation Time of the Proposed Algorithm} \label{apdx:comp_time}
In order to describe the computational time of the proposed approach, we run our algorithm over 50 bandwidth traces
and measure for different prediction window sizes ($W$) the average computation time for scheduling within one prediction window across all traces. We use a commodity MacBookPro laptop with 2.5GHz intel Core i5 processor and 4GB 1600MHz DDR3 RAM.
We vary $W$ from 1 to min($deadline(C)$, 10 minutes). We choose high prediction window sizes to show the computational efficiency of the algorithm for large problems.
As shown in in Fig.~\ref{fig : cpuTime},
our algorithm incurs very low computational overhead due to its linear nature.
Even for a window size of 10 minutes, the running time is only about 92 ms. The low overhead makes it feasible for the algorithm to run on low-end mobile devices.
Also, as expected, when we increase $W$, the running time almost increases linearly, indicating the scalability of our algorithm.


\end{document}